\documentclass{aa}

\usepackage[varg]{txfonts}

\usepackage{color}
\begin{document}
\title{Dust Impact Monitor (SESAME-DIM) Measurements at Comet 67P/Churyumov-Gerasimenko}



\author{Harald~Kr\"uger\inst{1}\thanks{\emph{Contact address:}
krueger@mps.mpg.de} 
\and  Klaus~J. Seidensticker \inst{2}
\and Hans-Herbert Fischer\inst{3}
\and  Thomas Albin \inst{1,4}
\and  Istvan Apathy \inst{5}
\and  Walter Arnold \inst{6,7}
\and  Alberto Flandes \inst{8,1}
\and  Attila Hirn \inst{5,1}
\and  Masanori Kobayashi \inst{9}
\and  Alexander Loose \inst{1}
\and  Attila P{\'e}ter \inst{5}
\and  Morris Podolak \inst{10}
}

\institute{Max-Planck-Institut f\"ur Sonnensystemforschung,     
              Justus-von-Liebig-Weg 3, 37077 G\"ottingen, Germany
  \and Deutsches Zentrum f\"ur Luft- und Raumfahrt, Institut f\"ur Planetenforschung, Rutherfordstra{\ss}e 2, 12489 Berlin, Germany
  \and Deutsches Zentrum f\"ur Luft- und Raumfahrt, 
  MUSC, Linder H\"ohe, 51147 K\"oln, Germany
  \and Medical Radiation Physics, Faculty VI, Carl von Ossietzky University, Oldenburg, Germany
   \and MTA Centre for Energy Research, Hungarian Academy of Sciences, 1121 Budapest, Hungary
   \and Department of Material Science and Material Technology, Saarland University, 66123 Saarbr\"ucken, Germany
 \and 1. Physikalisches Institut, Universit\"at G\"ottingen, Friedrich-Hund-Platz 1, 37077 G\"ottingen, Germany
\and Ciencias Espaciales, Instituto de Geof\'isica, Universidad Nacional Aut\'onoma de M\'exico, Coyoac\'an 04510, M\'exico, D.F. 
\and Planetary Exploration Research Center, Chiba Institute of Technology, Narashino, Chiba 275-0016, Japan 
\and Department of Geosciences, Tel Aviv University, Tel Aviv 69978, Israel
} 

\date{Received / Accepted }

\abstract 
{{\sf \em Context.} The Rosetta lander Philae successfully landed on the nucleus  of comet 67P/Churyumov-Gerasimenko 
on 12 November 2014.
Philae carries the Dust Impact Monitor 
(DIM) on board, which is part of the Surface Electric Sounding and Acoustic Monitoring Experiment (SESAME).
DIM employs piezoelectric PZT sensors to detect impacts by sub-millimeter and millimeter-sized ice and dust particles 
that are emitted from the nucleus and transported into the cometary coma. 

{\sf \em Aims.} The DIM sensor measures
dynamical data like flux and the directionality of the impacting particles. Mass and speed of the  particles
can be constrained  assuming density and elastic particle properties.

{\sf \em Methods.} DIM was operated during three mission phases of Philae at the comet: 
(1) Before Philae's separation from Rosetta at distances of about
9.6~km, 11.8~km, and 25.3~km  from the nucleus barycenter. In this mission phase 
particles released from the nucleus on radial trajectories remained undetectable because of significant obscuration by  
the structures of Rosetta, and no dust particles were indeed detected.
(2) During Philae's descent to its nominal landing site Agilkia, 
DIM detected one approximately millimeter-sized particle 
at a distance of 5.0~km from the nucleus' barycenter, corresponding to an altitude of 2.4~km 
from the surface. This is the closest ever dust detection at 
a cometary nucleus by a dedicated in-situ dust detector.
(3) At Philae's final 
landing site, Abydos, DIM detected  
no dust impact which may be  
due to low cometary activity in the vicinity of Philae, or due to shading by obstacles close to Philae, or both.

{\sf \em Results.} Laboratory calibration experiments showed that the material properties of the detected  particle are compatible 
with a porous 
 particle having a bulk density of approximately $\mathrm{250\,kg\,m^{-3}}$.  
 The particle could have been lifted off from the comet's surface by sublimating water ice. 
}


\titlerunning{Dust Impact Monitor (SESAME-DIM)}

\maketitle

\bibliographystyle{aa}

\section{Introduction}

\label{sec_introduction}

On 6 August 2014 the European spacecraft Rosetta \citep{glassmeier2007} encountered its target comet 
67P/Churyumov-Gerasimenko (hereafter 67P) and became the first artificial satellite of a cometary nucleus. 
Rosetta carried the lander spacecraft Philae on board \citep{bibring2007}, which on 12 November 2014 
successfully landed on the surface 
of 67P, at a heliocentric distance of 2.99~AU \citep{biele2015}. One of the scientific instruments on board Philae is the 
Dust Impact Monitor (DIM) which is part of the 
Surface Electric Sounding and Acoustic Monitoring Experiment  \citep[SESAME;][]{seidensticker2007}. 

Comet 67P is a short period comet with a present orbital period of about 6.6 years and aphelion and perihelion 
distances at 
 5.68~AU and 1.24~AU,
respectively. The knowledge about 67P before the arrival of Rosetta was
summarised by \citet{lamy2007}.
Nucleus images recently taken by the cameras on board Rosetta show an irregularly shaped body with a size of about 4~km
that consists of two lobes connected by a short depression, and a wide diversity in surface morphology \citep{thomas2015}.

In-situ investigations of comets 1P/Halley, 81P/Wild~2 and 9P/Tempel~1 showed 
that micron-sized {\em particle}s dominate the dust population in their comae within several hundred kilometers 
from their nuclei \citep{mcdonnell1991,tuzzolino2004,economou2013}. The existence of larger millimeter to 
centimeter sized particles in cometary comae is also evidenced by meteor streams 
that can be attributed to individual comets \citep{jenniskens2006}. Particles up to tens 
of centimeters were reported from the flyby of the 
Deep Impact spacecraft at comet 103P/Hartley~2 \citep{ahearn2011}.

Earth based observations before Rosetta's arrival indicated that particles in 
the size range from $\mathrm{10\,\mu m}$ to 1~mm or -- possibly even larger -- may exist in the 
coma of 67P \citep{tozzi2011}. Millimeter-sized  particles  also exist in the dust trail \citep{agarwal2007b}, and  recent
investigations with  Rosetta  revealed 
big millimeter to centimeter sized particles  in the close vicinity of the nucleus of 67P \citep{rotundi2015}.
Although a significant fraction of these large particles move slower than the escape speed near the cometary nucleus,
escaping  particles in this size range may considerably contribute to the nucleus' total mass loss.

\begin{figure}
   \centering
\includegraphics[width=0.45\textwidth]{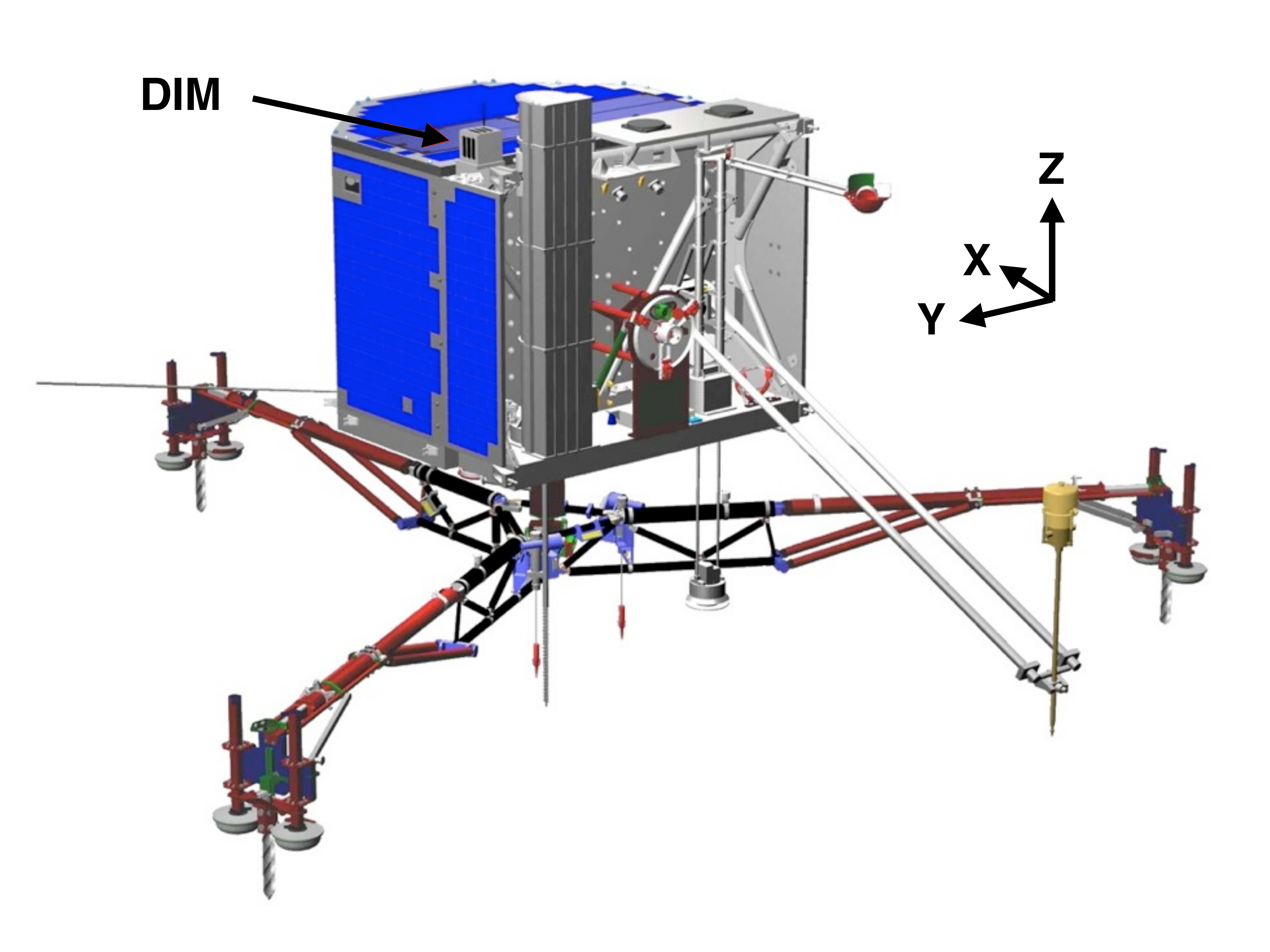}
\caption[]{
Rosetta lander Philae. The DIM sensor is visible at the top. 
Drawing ESA/ATG medialab.
}
\label{fig_philae}
\end{figure}

 The DIM sensor
is mounted on the top side of Philae 
 (Figure~\ref{fig_philae})
and measures sub-millimeter and millimeter sized particles hitting 
piezoelectric sensor plates. It measures
dynamical data like flux and the directionality of the impacting particles. Mass and speed of the  particles
can be constrained 
 assuming
particle density and elastic properties \citep{seidensticker2007}.

DIM was operated for a total of 280~minutes before the separation of Philae from Rosetta, for a total of 
35~minutes after the separation
when Philae was descending to its nominal landing site on  
the surface of 67P named "Agilkia", and for 280~minutes at the final landing site "Abydos". 
The goal of the DIM measurements at 67P was the determination of the abundance and the  particle size distribution 
in the submillimeter to millimeter size range to constrain the mass loss from the comet in the form of large dust  particles.
Measurements of the temporal variability in the dust flux were foreseen, aiming at  diurnal and seasonal 
variations in dust release from the cometary surface close to Philae's landing site. Furthermore, the material
properties of the detected  particles were to be constrained with DIM.

In Section~\ref{sec_rosetta} we give a brief overview of Philae's mission to comet 67P. 
Section~\ref{sec_dim}  describes the DIM instrument and its operation on board Philae, and in  
Section~\ref{sec_measurements} we present the DIM dust measurements at comet 67P. In Section~\ref{sec_analysis} 
we analyse and in Section~\ref{sec_discussion} we discuss our results.
Finally, in Section~\ref{sec_conclusions} we summarise our conclusions.

\section{Philae mission}

\label{sec_rosetta}

The Rosetta spacecraft was launched on 2 March 2004. After swing-bys at Earth and Mars, and flybys at
the asteroids \u{S}teins and Lutetia, Rosetta arrived at its target comet 67P in mid 2014. The spacecraft was brought 
into an orbit 
about the comet nucleus on 6 August 2014, with an initial altitude of 100~km. Until mid October 2014 the orbit was
stepwise reduced to 10~km.


Rosetta carried the lander spacecraft Philae on board which attempted the first ever landing on a cometary nucleus. 
Philae was separated from Rosetta on 12 November 2014 at 08:35:00~UTC. At that time Rosetta was at a distance of
22.7~km from the comet's barycenter, corresponding to an altitude of about 20.5~km from the surface \citep{biele2015}.
During the descent, Philae was rotating about its +Z axis. After seven hours descent, 
Philae reached the comet surface and made its first touchdown at 15:34:04~UTC, 
at the Agilkia landing site. The touchdown speed 
relative to the comet surface was $\mathrm{1.0~m\,s^{-1}}$. 

Due to failures of its 
anchoring mechanism and active descent system, Philae did not come to rest at  Agilkia. Instead, the 
spacecraft bounced and continued its journey across the surface of 67P. It came to its final rest about 
two hours later, at 17:31:17~UTC \citep{biele2015}. The final landing site, Abydos, is approximately 1.2~km 
away from Agilkia, and, to date, Philae's location is only known with
an uncertainty of approximately  21 x 34~$\mathrm{m^2}$ \citep{kofman2015}. 

At Abydos, Philae encountered very low ambient temperatures varying from 
 90~K to 130~K between night and day \citep{spohn2015}.
The insolation of Philae's solar arrays lasted only for about 1.5~hours per comet day \citep[comet rotation period 
$12.4043 \pm 0.0007$ h after its 2009 perihelion;][]{mottola2014}. 
The low temperatures, the short solar insolation period, as well as images of Philae's environment taken by
the cameras on board \citep{bibring2015},  which show ''cliffs`` and walls within a few meters distance, indicate that Philae came to rest in some type of cavity. 
Philae may be tilted by
approximately $90^{\circ}$, with its instrument balcony (-X side) pointing to the ground 
\citep[][{\em cf.} Figure~\ref{fig_philae}]{bibring2015}. 

Philae was successfully operated from battery power during its nominal mission at Abydos which lasted 
for about 56~hours. All instruments on board were operated during this period, including DIM. When the batteries
were exhausted, Philae went into hibernation because the insolation of the solar arrays was too low for a
continuing operation of the spacecraft.  During the following months, the comet was approaching the 
Sun and passed through its perihelion on 13 August 2015, leading to a gradual increase of the insolation. 
On 13 June 2015 contact with Philae was established again, opening the possibility for extended 
measurements at the comet surface that were never foreseen for Philae, i.e. during the highest activity 
phase of 67P. 

\section{Dust Impact Monitor}

\label{sec_dim}

The piezoelectric sensors of DIM are mounted on a cube with about 7~cm side length. Three sides of the  
cube are covered with  sensors,  
the other three sides are either closed by aluminum plates or left open for harness access. 
The DIM cube is mounted on the top side of Philae, and the three active 
sensor sides point in the +X, +Y and +Z directions in the Philae coordinate system (Figure~\ref{fig_dim}).
Each of the active DIM sides is divided into three equally sized segments that 
carry rectangular piezoelectric sensors made of PNZT7700 (Pb, Ni, Zi, Ti  -- hereafter PZT). 
The size of each segment is
$\mathrm{50\,x\,16\,x\,1\,mm^3}$. Adding all nine active segments leads to about $\mathrm{70\,cm^2}$
total sensor area. More details about the DIM sensor can be found in \citet{seidensticker2007} and 
\citet{flandes2013,flandes2014}.


Each particle impact onto one of the piezoelectric sensor plates generates an electric pulse that is 
registered with the instrument electronics. The output signal can be approximated by a 
damped sine wave, and the measured signals are analysed by 
Hertz' theory of contact mechanics \citep{hertz1882,johnson1985}.  The
amplitude and the width of the first half-sine pulse are 
used to derive particle properties like the reduced modulus and the mass. To calibrate the instrument signals, 
we performed a large number of drop experiments with particles made of different materials, including water
ice as a cometary analogue material \citep{flandes2013,flandes2014}.



\subsection{DIM impact measuring values}

 Two parameters 
-- the impact duration $T_c$ and the peak voltage $U_m$ -- are derived from the signal curve by the
instrument electronics. These values are transmitted to Earth \citep{flandes2013}. 
 
To distinguish noise or false events from real dust impacts, we use the term {\em impact} for true dust
impacts only, and we use the term {\em noise}, {\em false} or {\em long event} for all other registered events
({\em cf.} Section~\ref{sec_impacts} for a definition of these event types). The simple 
term {\em event} is used in a general sense, referring to both real dust impacts and all other types of 
registered events. In addition, we use the term {\em false signal} for events that are characterised by
the instrument software as true dust impacts but for which we know by other means that they cannot be due to 
dust impacts. The instrument software is not able to distinguish false signals and dust impacts. 
 We use 
the term "dust" as a synonym for both ice or dust particles. 

\subsubsection{Impact duration}

The impact duration is derived from the number of counts of a 20 MHz clock.
The conversion between impact duration on a logarithmic scale to microseconds is given by Equation~\ref{eq_dim1}
in Appendix~1.

So-called "false" and "long" events are identified by using algorithms 
specified by \citet{peter2001}, see also Section~\ref{sec_impacts}. These events are defined according to the
measured impact duration.
They are counted with an onboard
counter and only their total numbers are transmitted to Earth. The impact durations are transmitted for true dust 
impacts, but neither for false nor for long events.  

\subsubsection{Peak voltage} 

The instrument provides the peak voltage of a single dust impact after 
logarithmic amplification. The amplifier transfer function of the DIM flight model is
given by Equation~\ref{equ_transfer} in Appendix~1, and for storage a fixed logarithmic scaling is used 
as given by Equation~\ref{eq_dim2}. 

 The transfer characteristics of the logarithmic DIM amplifier 
can vary with time. It was regularly checked by a calibration procedure ({\em cf.} Section~\ref{sec_commanding}). 

\subsubsection{Event time}
In addition to the impact duration and the peak voltage of a signal, the time in UTC when the event occurred 
is also stored for up to 350 events transmitted in Burst Continuous Test2 mode (not in Burst Continuous mode; see Section~\ref{sec_commanding} for a description of these
measurement modes). 

\subsection{DIM operational modes}

\label{sec_commanding}

DIM can be operated by telecommands from the Earth. Before any DIM 
 measurement is started, a few operations have to be performed to guarantee that DIM is properly working. Here, 
we can only give an overview, more details about the DIM operation can be found in
\citet{fischer2012}.

After DIM is switched on: 
(A) a power check is performed to verify that the supply voltages are 
within predefined limits.  Then 
(B) electronic noise is measured on the DIM amplifier 
 (with the DIM sensor being disconnected): 
Starting from a very low value, the detection threshold (so-called margin, {\em cf.} Section~\ref{sec_margin}) 
is increased in steps of 10~dB until no false event is detected anymore. The rates of false events typically 
encountered in flight were
such that the margin was set to 30~dB or 40~dB ({\em cf.} Tables~\ref{tab_operation} and
\ref{tab_operation2}). This means that a true dust impact must have an amplitude of at least  approximately 
0.1\,mV or 0.25\,mV, respectively, 
to overcome the threshold of the amplifier \citep{krueger2012}. In a next step 
(C), a DIM sensor test checks if all three sensor 
sides are operational. An electrical pulse 
(approximately 5 V for $\mathrm{10\,\mu s}$) 
is applied to each sensor side, and the response is registered in the same way 
a dust impact would be measured. Finally (D), the electronics performs a DIM calibration to re-calibrate the transfer 
characteristic of the logarithmic amplifier and to check the time measuring ($T_c$) circuit: Two 
test pulses are applied to the logarithmic amplifier. 
Pulse height and duration are $\mathrm{1\,mV}$, $\mathrm{8\,\mu s}$ for low-level and 
$\mathrm{100\,mV, 20\,\mu s}$ for high-level, respectively. 

The results of the calibration procedure 
were supposed to be used to re-calibrate the amplifier transfer function in the data evaluation on Earth. 
However, it turned out after launch that the
high-level value is in saturation  for temperatures below $+20^{\circ}\,\mathrm{C}$ so that this re-calibration procedure could not be applied. 
 We therefore used the amplifier transfer function measured before launch throughout the data
evaluation presented in this paper, which is given by Equation~\ref{equ_transfer}. The uncertainty 
caused by this deficiency is taken into account in our error estimation (cf. Section~\ref{sec_temp})

After successful execution of the above listed steps, DIM is ready for measuring dust impacts. 
Two measurement modes were used to measure dust at the comet:

Single events on one sensor side can be registered in the so-called Burst Continuous (BC) Mode. The measured  
peak amplitude $U_m$ and the impact duration $T_c$ are stored in a compressed way: They are scaled 
to a logarithmic scale ${\bf U}_m$ and ${\bf T}_c$ according to Equations~\ref{eq_dim1} and \ref{eq_dim2}, and 
the counts for impacts  or false signals with a particular [${\bf U}_m$, ${\bf T}_c$] combination are stored 
in memory cells of different sizes, 
depending on the expected frequency of such events 
 (We use regular characters ${U}_m$ and ${ T}_c$ to distinguish the raw uncompressed data from the 
compressed logarithmic values which we denote with bold characters ${\bf U}_m$ and $ {\bf T}_c$).
Each BC measurement starts with a 10~s instrument warm-up period which is 
not included in the measuring time.

In addition to the BC mode (which delivers the [${\bf U}_m, {\bf T}_c$] matrix), the instrument 
can be operated in a so-called Burst Continuous Test2 mode (BCT2). Similar to the BC mode, the BCT2 mode delivers  
$U_m$ and $T_c$ for each individual impact. Here the raw uncompressed data are transmitted instead of the
compressed logarithmic values, and in addition 
the event time when the impact or false signal was registered. 
The number of data sets for impacts, which can be stored and transmitted to Earth in BCT2 mode, is limited to 350.

\subsection{Measurement of single impacts}

\label{sec_impacts}



A DIM measurement consists of a series of single event measurements. Each of these 
measurements detects and evaluates one event by comparing the amplified sensor signal $U_{\mathrm{out}}$ 
with the detection threshold voltage 
$U_{\mathrm{thr}}$ ({\em cf.} Section~\ref{sec_margin}). The underlying assumption is that the initial part of the dust impact signal can be approximated by the first half of a sine wave. An impact should thus show up as a voltage, 
crossing the threshold voltage upwards, followed by a second threshold crossing downwards. The period between the two threshold  crossings defines the impact duration $T_c$.

If the amplified sensor signal crosses the detection threshold too early (less than 1 ms after the single event measurement was initialized), the event is classified as a "false event", else it is accepted as the beginning of a potential real impact. 
 The measurement
is stopped after 0.5~ms. The event is ruled out as a "long
event", if no second threshold crossing was detected during the
measurement duration. A true dust impact is thus characterised
by its occurrence later than 1 ms after the initialization of the
measurement and a duration of less than 0.5~ms.

Regardless of the event type (dust impact, false or long
event), an adjustable dead time, the so-called sensor signal decay
time \citep{flandes2013}, is included after the end of each single
event measurement.  It was set to 5 ms for all measurements. 
Apart from the signal decay time, several
waiting and latency periods are added to the total duration of a
single impact measurement. All latency periods add up to a total
instrument dead time of approximately 10~ms \citep{fischer2014}.


\subsection{Margin}


\label{sec_margin}


The DIM electronics can detect an event if the amplified signal voltage exceeds an adaptive threshold value $U_{\mathrm{thr}}$. 
The detection threshold $U_{\mathrm{thr}}$ is the sum of an adjustable margin and a signal average. 
The margin can be increased 
in steps of 10~dB in the range from 10 to 70~dB. Each step changes the threshold voltage by approximately 0.3~V. 
The signal average is determined by the DIM electronics with a time constant of approximately $1~\mathrm{s}$. It varies 
slowly with impact properties and frequency, aiming at covering a wide range of event voltage levels. 
Measuring values for the respective sensor side of the signal average are transmitted to Earth if sampling of 
average values is commanded.

\subsection{Electronics ageing}

\label{sec_ageing}

The performance of the DIM instrument electronics has been regularly checked since Rosetta's launch in 2004. 
So-called payload checkouts of the instruments on board were performed approximately every six months between 2004 and 2011,
 and again in 2014. No such procedures were executed between 2011 and  early 2014 when 
Rosetta was in hibernation.  

During these checkouts the DIM operational procedures described in Section~\ref{sec_commanding} were executed. 
They did not reveal any ageing of the instrument electronics,  except a deficiency in the calibration 
procedure (D) described in Section~\ref{sec_commanding}. We therefore conclude that the
performance of the DIM electronics remained stable during the entire Rosetta mission.

\subsection{Temperature dependence}

\label{sec_temp}
 
Operational periods when the DIM electronics noise was measured and sensor tests   were
performed during flight showed no significant temperature dependence (operational procedures B and C in
Section~\ref{sec_commanding}). On the other hand, a strong correlation 
between calibration measurements (operational procedure D in
Section~\ref{sec_commanding}) and the estimated DIM electronics temperatures could be observed: Between $-40^{\circ}\mathrm{C}$ and 
$+30^{\circ}\,\mathrm{C}$ the impact durations for the high calibration peaks 
decreased by 8\%, and those of the low calibration peak 
by 20\%. The low calibration peak amplitudes varied within $\pm4\%$ and showed a non-monotonous dependence as a function of the
electronics temperature. 
 High calibration peak amplitudes were not included in the analysis since the values were 
 saturated in a wide temperature range.
The temperature dependence of the system, to a large extent, can be attributed to the 
temperature dependence of the logarithmic amplifier.


\section{Dust measurements at comet 67P}

Before Rosetta's launch, DIM was designed to measure dust particles after Philae has 
landed on the surface of the
cometary nucleus. Measurements before the separation from Rosetta and during Philae's descent
were not foreseen at the beginning, and they were only  added later in the planning process 
of the mission. 

\label{sec_measurements}

\subsection{DIM operation}

\label{sec_operation}

After Rosetta arrived at comet 67P on 6 August 2014 and before Philae was released on 12 November 2014,
DIM was operated three times, namely on 16 and 17 October 2014 and 
on 12 November 2014 (Tables~\ref{tab_operation} and \ref{tab_operation2}, measurements 1 to 29). 
When Philae was connected to 
Rosetta, the DIM sensor was strongly shielded by the
Rosetta structures. Only the DIM +X sensor side was operated in these time intervals
because it was the least obscured side (Section~\ref{sec_geometry}).
DIM was operated in BCT2 mode during a total of 280~minutes. No dust impact and only very few 
false signals were registered. 

\begin{figure}
   \centering
\includegraphics[width=0.49\textwidth]{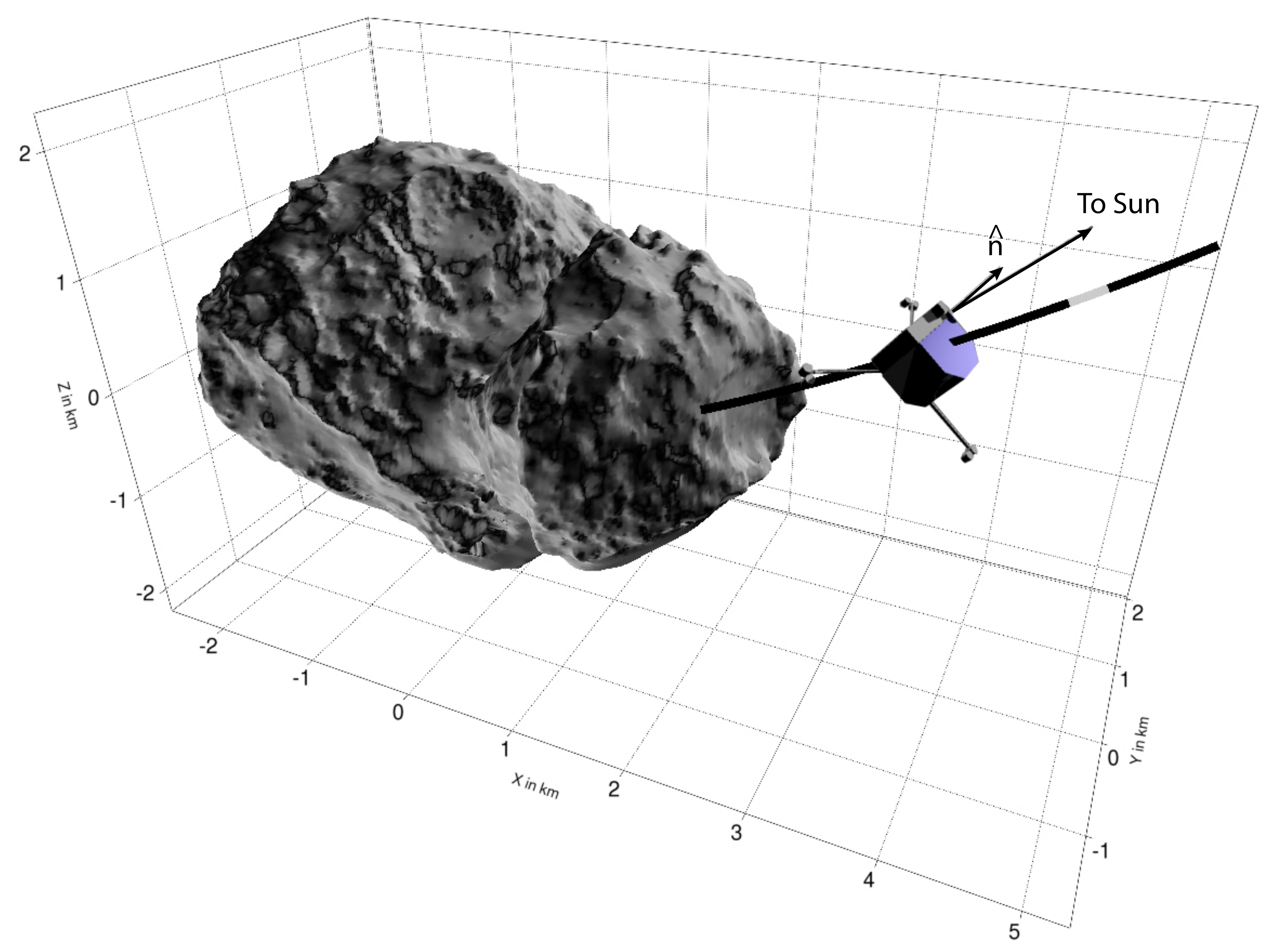}
\caption[]{
Philae trajectory close to the surface of comet 67P on 12 November 2014. The trajectory 
is shown in the comet fixed coordinate system. 
The position of Philae at the time of the single particle detection is illustrated. 
The vector $\hat{n}$ represents the normal to the +Y side of DIM. 
For the nucleus shape we used the 
ESA Engineering Model CSHP\_DV\_096\_01\_\_00161.ROS.
}
\label{fig_traj}
\end{figure}

\begin{figure}
   \centering
   \hspace{-10mm}
\includegraphics[width=0.54\textwidth]{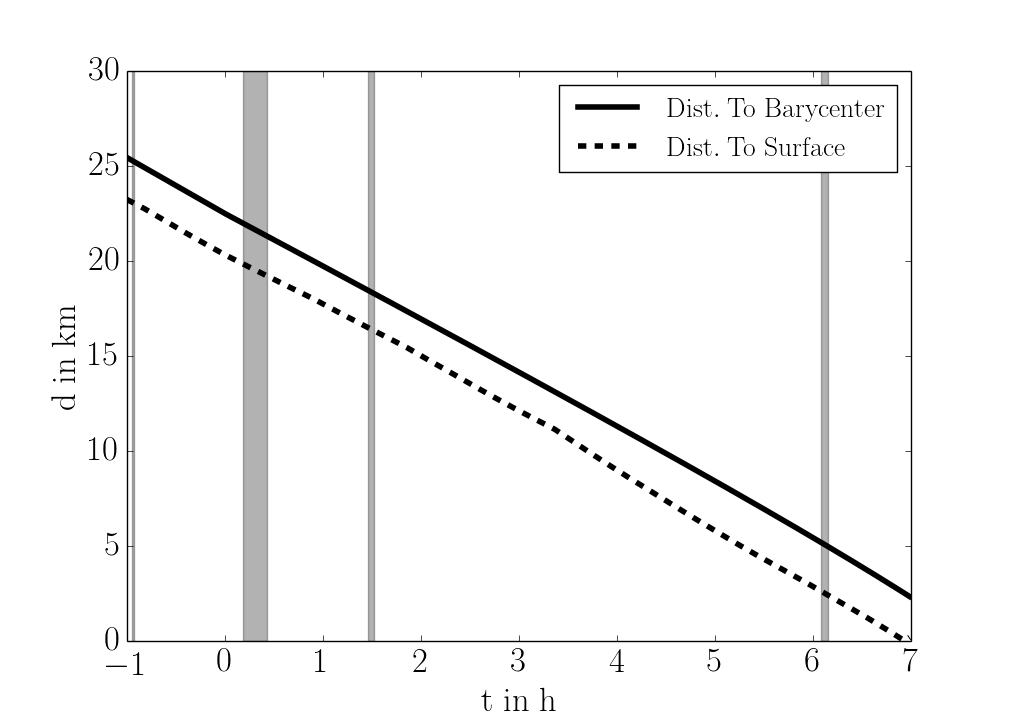}
\caption[]{
Philae's altitude above the nucleus surface and distance from the comet's barycenter, respectively, 
as a funtion of time during the descent to the nucleus of comet 67P. The time
$t=0$ refers to Philae's separation from Rosetta on 12 November 2014; 08:35:00~UTC. 
Grey vertical bars indicate the operational periods of DIM. 
}
\label{fig_descent}
\end{figure}

When Philae was descending to its first touchdown point Agilkia on 12 November 2014, DIM was operated in 
BCT2 mode during a total of 35~minutes. Measurements were taken sequentially with all three sensor sides 
(Table~\ref{tab_operation2}, measurements 30 to 41). Philae's descent trajectory close to the nucleus
and the altitudes of 
the DIM measurement periods are shown in Figures~\ref{fig_traj} and \ref{fig_descent}.

Before Philae landed on 67P the entire operational procedures performed on board during the
descent were tested with the Philae Ground Reference Model (GRM) at Deutsches Zentrum f\"ur Luft- und
Raumfahrt (DLR) in Cologne, which is a twin of the Philae spacecraft  in space. During these tests a 
high rate of false signals was registered 
with the DIM electronics. 
This noise occurred only when insolation of the solar arrays was simulated, i.e. the signals were due to 
cross-talk \citep{hirn2015a}.
Therefore, we expected a similarly high rate of false  signals 
for the measurements with the DIM flight model during Philae's descent to the 
nucleus surface on 12 November 2014. 

To suppress the occurrence of such signals, we operated the instrument with  an increased margin 
setting for a few of the measurements during the 
descent: One measurement sequence was performed with
margin 40~dB from 08:38~UTC to 08:46~UTC, while this sequence was repeated with a margin of 50~dB from 08:50~UTC 
to 08:57~UTC. A high rate of false signals was indeed registered during a few of the measurement periods 
with the lower margin
setting, while the higher margin effectively prevented the occurrence of false signals. The higher 
margin setting raised the detection threshold of the instrument to somewhat larger particles ({\em cf.} 
Section~\ref{sec_sensitivity}) so that only particles with impact signals in excess of about 1~mV could be 
measured. As during the GRM tests performed earlier, the false signals were likely caused by Philae's solar energy 
generators ({\em cf.} Section~\ref{sec_events}). 

At the final landing site, Abydos, DIM was operated in BC mode for a total of 280 minutes on 13 and 14 
November 2014. 
Here, again all three sensor sides were operated sequentially. 
Only few false signals were recorded (Table~\ref{tab_operation3}, measurements 42 to 71). 

\subsection{Dust impact identification}

\label{sec_events}

Reliable identification of false signals in the data set is a prerequisite for the analysis of
dust at 67P based on the DIM measurements. In order to characterise the   
behaviour of the instrument with respect to false signals, we analysed DIM data obtained with both the  
 Philae GRM, as well as the data from the DIM Flight Model (FM) on board Philae. 

\subsubsection{Signal distribution in the [$U_m, T_c$] diagram}

The measured impact durations $T_c$ and signal amplitudes $U_m$ can be used to characterise  false
events in the DIM data set. It turned out that the  false signals recorded with  
the GRM instrument during ground tests show a  behaviour  very similar to those measured with the flight 
instrument on board Philae. Therefore, we concentrate our analysis on the flight data in the following.

\begin{figure}
   \centering
   \hspace{-7mm}
\includegraphics[width=0.5\textwidth]{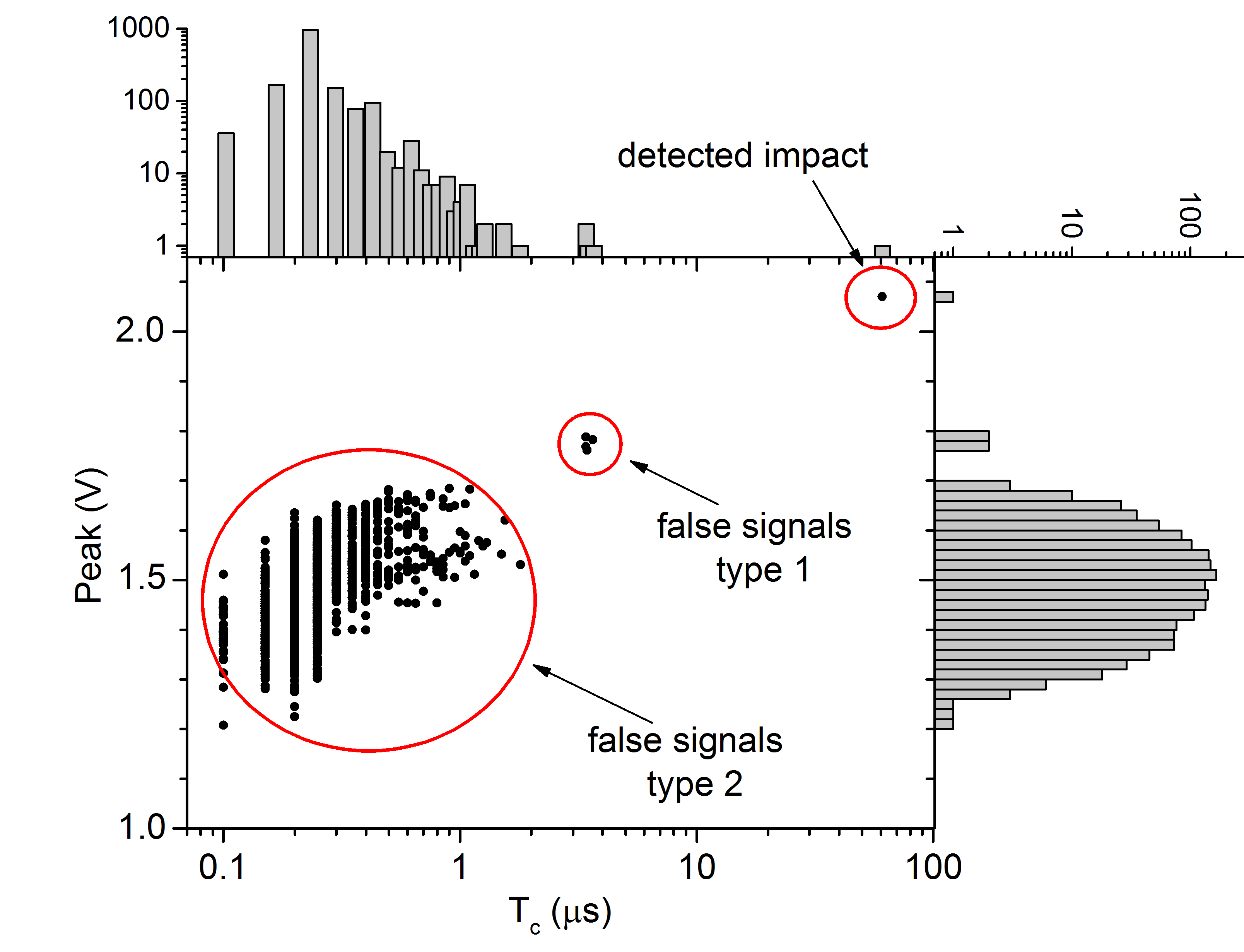}
\caption[]{
All detected events (false signals plus dust impacts) in the [$U_m,T_c$] diagram.
Vertical and horizontal histograms indicate the numbers of detected events. 
}
\label{fig_noise}
\end{figure}

We show the distribution of the measured signals in the [$U_m, T_c$] 
diagram in Figure~\ref{fig_noise}. The false signals cluster in a very limited region in the [$U_m, T_c$] plane. They mostly occur with 
very short impact durations up to about $\mathrm{1.5\,\mu s}$, and amplified signal amplitudes  
$ 1300\,\mathrm{mV} \lesssim U_{out}  \lesssim 1700\,\mathrm{mV} $.
A second group of false signals showed impact durations of approximately $\mathrm{3.4\,\mu s}$ 
and signal amplitudes $U_{out} \approx \mathrm{1800~mV}$. Clearly
separated from this region we detected one single event with a contact 
duration of $\mathrm{61~\mu s}$, and a signal amplitude of 2070~mV. 

We never detected false signals with such a large impact duration, neither during the GRM tests nor 
in earlier FM data. Laboratory experiments performed with ice spheres of about 0.6~mm radius yielded
impact durations below $6\,\mu \mathrm{s}$ \citep{flandes2014}.
However, such a large impact duration could be obtained during 
laboratory experiments with porous particles ({\em cf.} Section~\ref{sec_properties}). We therefore interpret this 
event as a true dust impact. It was registered with the DIM +Y sensor side on 12 November 2014, 14:43:47~UTC
during measurement number 40.


\subsubsection{Temporal behaviour of false signals}

During some measurement periods when Philae was descending to the nucleus surface, false signals 
occurred with a relatively high frequency up to $\mathrm{110\,s^{-1}}$ (Table~\ref{tab_operation2}). 
Very similar signal frequencies were also recorded during tests with the Philae GRM due to cross-talk. 

With such high rates of false signals one may ask whether DIM was able to detect any dust impacts in between
the individual false signals. Here, a consideration of the dead time of DIM gives some insights. As described in
Section~\ref{sec_impacts}, 
as a  good approximation we consider the dead time 
to be 10~ms for both the detected and the false signals. Given that the total event rates during periods with the
highest number of false signals are in the range $\mathrm{30\,s^{-1}}$ to 
$\mathrm{110\,s^{-1}}$,
the detection of  real impacts during these periods was extremely unlikely. We conclude that DIM was
not able to detect any dust particles during these noisy periods. 
Hence, for estimations of dust  fluxes \citep{hirn2015b}, the use of the length of the ''quiet`` 
period when no false signal was recorded is preferable over the total length of the BCT2 measurement.

It should be noted that the single dust impact recorded on 12 November 2014, 14:43:47~UTC was measured 
during a BCT2 measurement with 100~s measurement time (Table~\ref{tab_operation2}, measurement 40). False signals were recorded only for 2~s after the
start of this measurement, and the dust impact was measured 80~s later. No false signal was recorded in between.
This clearly separates the dust impact from the false signals recorded in this measurement interval.

\subsubsection{False events due to cosmic radiation}

A potential source of false events can be hits by high energy cosmic ray particles since
PZTs show some sensitivity to high energy radiation. Two major components of cosmic radiation can be
distinguished: galactic cosmic rays (GCRs) originating from outside the solar system and solar cosmic radiation. 
GCRs consist of energetic charged particles (86\% protons, 12\% alpha particles, and heavier ions and electrons typically
in the energy range of 1~MeV to $10^{14}\,\mathrm{MeV}$). Solar cosmic radiation is composed of charged particles 
having a softer (eV to GeV) energy spectrum than GCRs. Most of the time the GCR component dominates, albeit
during solar flares the flux of the solar cosmic radiation can be orders of magnitude higher. 

 High energy radiation hitting the 
PZT can generate acoustic signals similar to mechanical impacts that could mimic a dust particle impact. 
Laboratory experiments with PZT ceramics showed that only bursts of penetrating ions, neutrons, 
or gamma-photons 
may cause a signal with a strength of approximately 1~mV \citep{miyachi2006,miyachi2010,adline1985,holbert2005}. 
No single event was detected either with high energy 
ions or high energy photons, i.e. gamma rays. We can therefore exclude cosmic radiation as a source of 
false signals in our DIM measurements. 

\subsection{Detection geometry}

\label{sec_geometry}

\begin{figure}
   \centering
   \hspace{-2mm}
\includegraphics[width=0.5\textwidth]{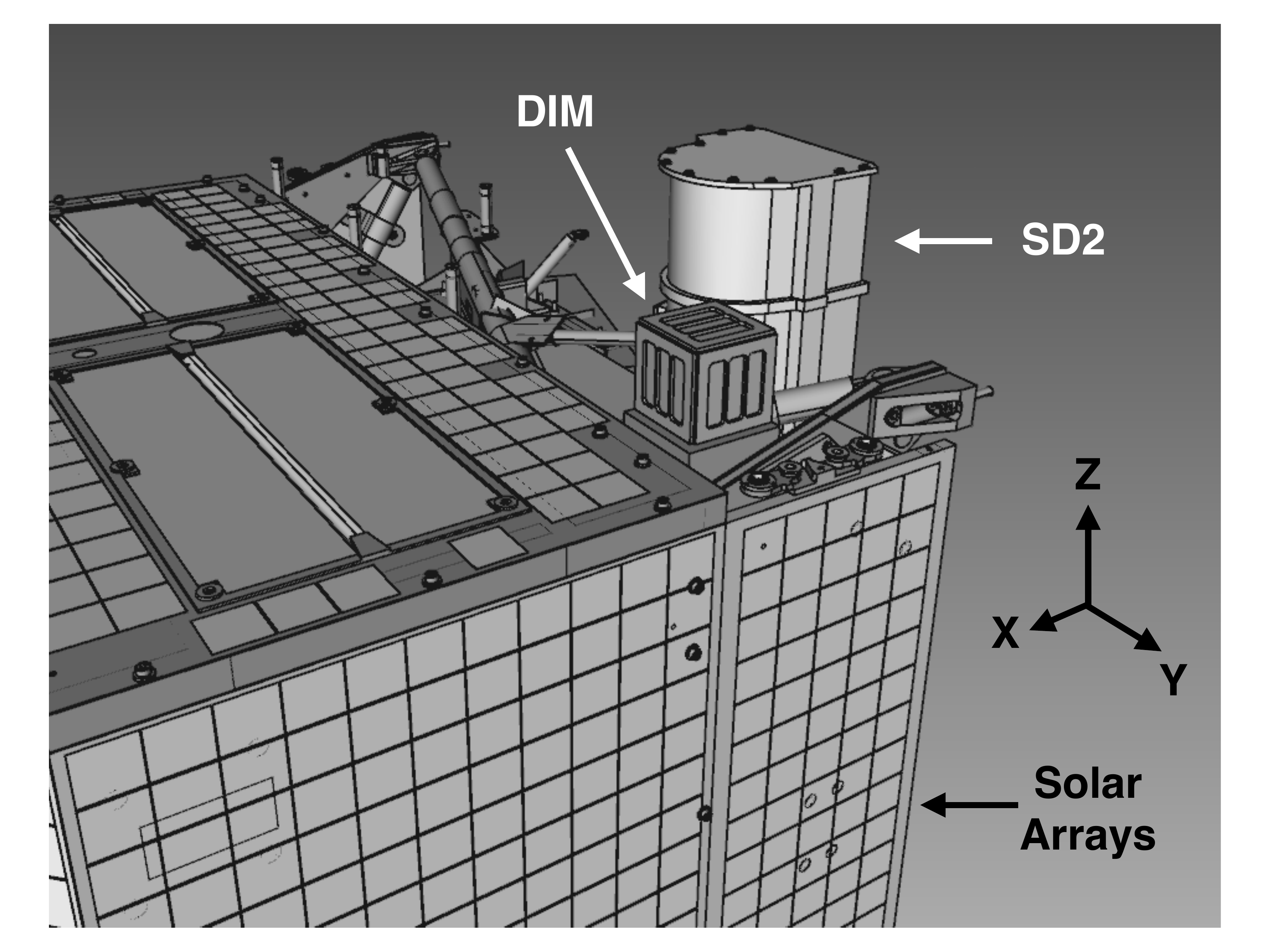}
\vspace{-2mm}
\caption[]{
Detail of Philae. The DIM cube with its three active sensor sides is visible at the top.
The housing of the SD2 drill is also indicated.
}
\label{fig_dim}
\end{figure}

 Before the separation of Philae from Rosetta, the DIM sensor was heavily 
obscured by structures of Rosetta and by
Philae ({\em cf.} Figure~\ref{fig_dim}). 
While the DIM +X sensor side was only partially obscured by overhanging
structures of Rosetta,  the +Y sensor was almost entirely shielded,
and the +Z sensor was completely covered. Therefore, only the 
+X sensor was operated before Philae's separation. 



After the separation from Rosetta the detection geometry was completely different. 
Figure~\ref{fig_dim} shows the DIM cube mounted on the top of Philae. 
DIM is partially shielded by Philae structures and the shielding differs for 
the three sensor sides. The +X side can only detect dust particles approaching from the +Z direction 
(top side of Philae) while Philae shields particles approaching from the --Z hemisphere entirely. The +Y side is  
partially shielded for particles approaching from the --Z hemisphere. 
Finally, the +Z side is  partially shielded by the housing of Philae's  SD2 drill 
(light grey structure in Figure~\ref{fig_dim}), but it has the largest 
field-of-view of the three sensor sides \citep{hirn2015b}.

After the separation from Rosetta, Philae was rotating about its Z axis so that
the +X and the +Y sides of DIM scanned a complete circle during one Philae revolution, with an 
offset of $90^{\circ}$ ({\em cf.} Figure~\ref{fig_dim}). 
Initially after the separation, Philae's rotation period was about 5~minutes. 
After the landing gear was unfolded the rotation period increased to about 8.6~min. The landing 
gear was unfolded at about 08:41~UTC (R. Roll, priv. comm.), i.e. during DIM 
measurement 30 (Table~\ref{tab_operation2}).



\section{Results and analysis}

\label{sec_analysis}


\subsection{Dust detections}

As described in Section~\ref{sec_operation}, before and after the separation of Philae from Rosetta, the 
operational conditions of the DIM instrument
differed. 
We therefore distinguish three measurement intervals and configurations: 
\begin{itemize}
\item[(1)] Philae connected to Rosetta, 
\item[(2)] Philae's descent to the comet surface (landing site Agilkia), 
\item[(3)] Philae at its final landing site Abydos. 
\end{itemize}

\subsubsection{Philae connected to Rosetta}

\label{sec_connected}

DIM measurements on 16 and 17 October 2014 were taken at a distance of 9.6~km and 11.8~km from the nucleus 
barycenter, respectively, when Rosetta was orbiting the comet nucleus on an eccentric 
trajectory. A similar measurement was performed on 12 November 2014, about one hour before Philae's 
separation from Rosetta, at a distance of 25.3~km 
(Meas. numbers 1 to 29; Tables~\ref{tab_operation} and \ref{tab_operation2}). In all three cases
only the +X sensor of DIM was measuring because the other sensor sides were strongly obscured 
(Section~\ref{sec_geometry}).
 
During these measurements the orientation of Rosetta was such that the DIM +X sensor was pointing within 
$\pm 10^{\circ}$ perpendicular to the comet's nadir 
direction and perpendicular to the direction of motion of Rosetta.
With this orientation the DIM +X sensor could neither detect particles approaching radially from the nucleus  
nor particles on circular orbits about the nucleus. On the other hand, particles on highly bent
ballistic trajectories or particles with highly eccentric orbits
could  have been detected by  the DIM +X side. No dust impact was detected in this configuration
during a total of about 280~minutes measurement time.

\subsubsection{Philae's descent to the comet surface}

Philae was separated from Rosetta on 12 November 2014 at 08:35:00 UTC, and during its 
seven hour descent to the nucleus surface, all three DIM sensor sides were operated repeatedly with 
a total measuring time of 35~minutes (Table~\ref{tab_operation2} and Figure~\ref{fig_descent}). 
During several of these periods the 
recorded rates of false signals were very high so that no dust impact
was detectable between the noise events because of dead time of the instrument electronics 
({\em cf.} Section~\ref{sec_measurements}). 

Only one single dust impact was detected  during the descent to the
nucleus surface before the
first touchdown at Agilkia. The particle was detected with the DIM +Y side at 14:43:47~UTC at
a distance of 5.0~km from the nucleus' barycenter, corresponding to an altitude from the nucleus 
surface of about 2.4~km.  At this time Philae
was at a latitude of $22^{\circ}$ and 
a longitude of $346^{\circ}$ in the comet fixed reference frame defined 
by \citet{preusker2015}, and the spacecraft's velocity vector was 
$[v_x, v_y, v_z] = [-0.817, -0.232, -0.434] $ (in $\mathrm{m\,s^{-1}}$). 
The location of the particle detection 
w.r.t. the nucleus is shown in Figure~\ref{fig_traj}.


In order to constrain the particle dynamics from the DIM measurement, the velocity
vector (i.e. impact direction and speed) has to be determined. Given that dust 
particles can hit each DIM sensor side from a large range of impact angles from almost half
a hemisphere, the impact direction is only little constrained  
(details of the DIM detection geometry will be discussed by \citet{hirn2015b}). In addition,
the speed measurement is connected with a large uncertainty as well \citep{flandes2013,flandes2014}.
It is therefore impossible to determine the particle dynamics with high accuracy
for a single detected particle. 

The derivation of the impact direction of the particle is further complicated 
because Philae was rotating about its Z axis during the descent. Therefore, the +X and +Y 
sensors of DIM scanned an entire circle during one spin revolution of Philae. 
No information about Philae's orientation during the descent was provided by the Philae Science Operations \& 
Navigation Center (CNES) in Toulouse/France at the time of this writing. Furthermore, attempts to determine
the orientation from the magnetic field measurements of the Rosetta Magnetometer and Plasma Monitor (ROMAP)
instrument on board Philae \citep{auster2015} led to inconclusive results for the interpretation of the DIM 
data. We  attempted to derive the orientation of the DIM sensors as a function of time from the 
output signals of Philae's solar arrays (T. Albin et al., manuscript in preparation). This 
analysis is still ongoing, and we therefore have to postpone a detailed discussion of the DIM detection 
geometry and the possible particle dynamics to a future publication.

\subsubsection{Final landing site Abydos}

All three sensor sides of DIM were operated  at the Abydos landing site for a total of 280~minutes 
(Table~\ref{tab_operation3}). Only very few false signals and no dust impact were recorded. 
The DIM health checks performed at Abydos did not indicate any non-nominal operation of DIM.

\subsection{Particle properties}


\label{sec_properties}

From the detected signal amplitude, $U_m$, and the impact duration, $T_c$, we can constrain the
radius, $R$, and the impact speed, $v$, of the particle \citep{seidensticker2007,flandes2013}:
\begin{equation}
T_c = 5.09 \,\left ( \frac{R^5 \,\rho^2}{v \,E_r^2} \right)^{1/5}       \label{equ_tc}
\end{equation}
and
\begin{equation}
U_m = \frac{3.03 \,d_{33} \,E_r^{0.4} \,\rho^{0.6} \, R^2 \, v^{1.2}}{C},       \label{equ_u}
\end{equation}
where $E_r$ is the combined reduced modulus of the PZT sensor, $E_{PZT}$ and the impinging particle $E_{p}$\citep[see][for details]{flandes2013}.

From Equations~\ref{equ_tc} and \ref{equ_u} we define  
\begin{equation}
a = \left (\frac{5.09}{T_c} \right)^5 \left (\frac{\rho}{E_r} \right )^2
\end{equation}
and 
\begin{equation}
b = \frac{3.03\,d_{33}\,E_r^{0.4}\,\rho^{0.6}}{U_m\,C} ,
\end{equation} 
and solve for $R$ and $v$, getting:  
\begin{equation}
v = \frac{a^{1/4}}{b^{5/8}}            \label{equ_v}
\end{equation}
for the impact speed, and
\begin{equation}
R = \left (\frac{1}{a^{6/5} \, b} \right )^{1/8}    \label{equ_r}
\end{equation}
for the radius of the particle.

The piezoelectric constant, $d_{33}$, the PZT's capacitance, $C$, and the reduced Young's modulus,
 $E_r$, are known \citep[see][]{flandes2013,flandes2014}. These constants depend, either on the 
properties of the PZTs or the impacting particles or both. $\rho$, is the particle's bulk density. A summary of these constants is given in Table~\ref{table:PIEZO}.

The $U_m$ and $T_c$ values are measured by the DIM electronics. However, as mentioned in 
Section~\ref{sec_dim}, 
only the amplified output voltage (or $U_{out}$) is stored and, therefore, $U_m$ needs to be derived from the known 
amplifier transfer function of the DIM electronics, which is given by Equation~\ref{equ_transfer} in Appendix~1. 


From our calibration experiments, it turned out that the $U_m$ and $T_c$ values recorded by the DIM electronics are sensitive to the shape of the signal produced by the particle impacts as well as on the margin set for the measurements. For the particle detected by DIM during the descent, $U_{out}=2070\,\mathrm{mV}$ and thus, from Equation~\ref{equ_transfer}, $U_{m}=2.45\,\mathrm{mV}$. Given the small input voltage and very long impact duration ($T_c=61\,\mu \mathrm{s}$), we assume that the impacting particle is quite likely a porous  conglomerate with reduced Young's modulus larger than $10\,\mathrm{MPa}$. This assumption is supported by calibration experiments performed with aerogel particles, having a bulk density of $\rho\approx 250\,\mathrm{kg\,m^{-3}}$ and Young's modulus $E_p=15\,\mathrm{MPa}$. The signal amplitudes and impact durations measured with aerogel
are consistent with the values recorded by the DIM electronics at the comet (see Appendix~2 for details). We therefore use 
the results from our experiments with aerogel particles to estimate the properties of the particle detected by DIM.

\begin{figure}
   \centering
\includegraphics[width=0.5\textwidth]{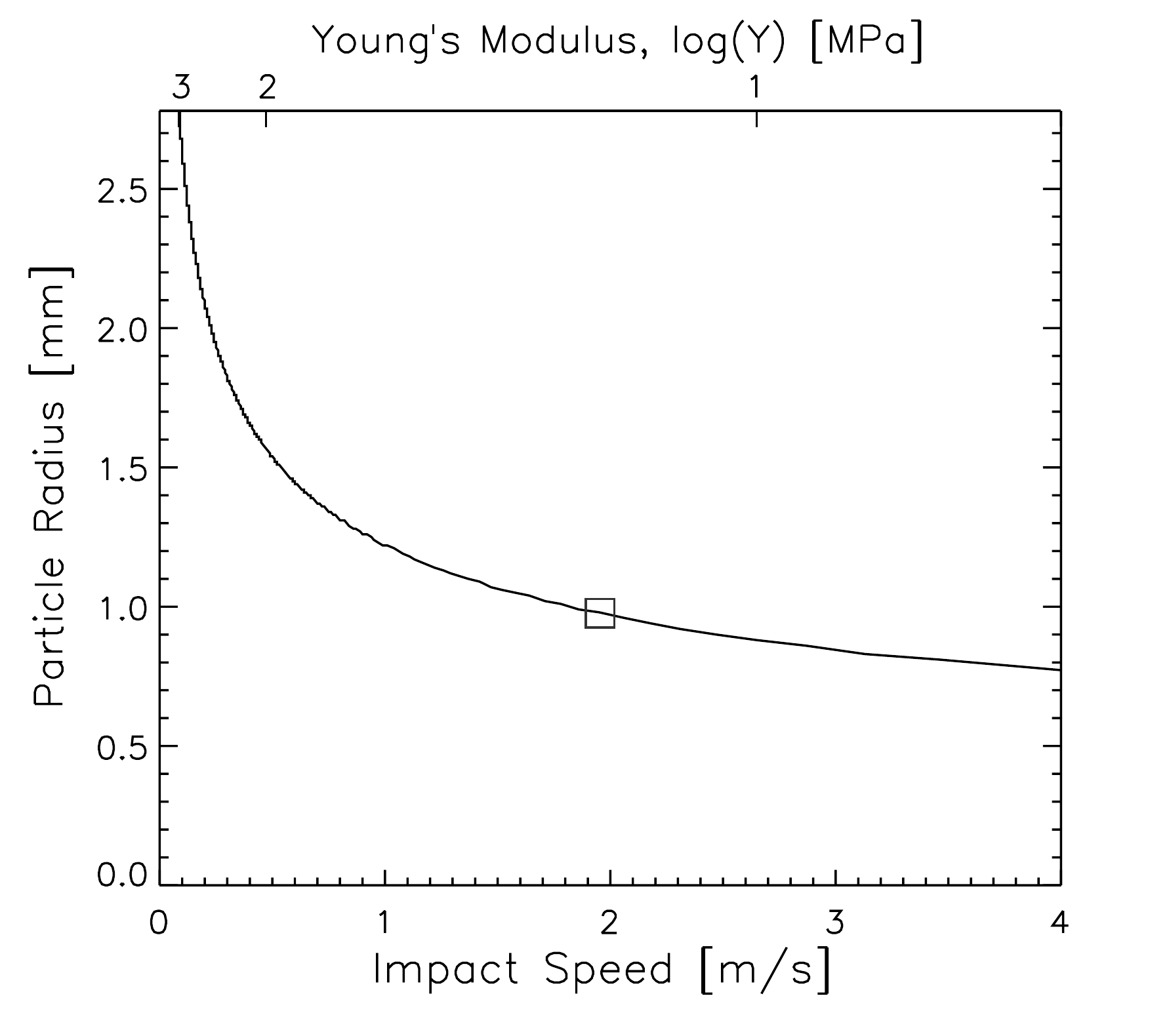}
\caption[]{Relationship between the impact speed, $v$, radius, $R$, and the Young's modulus, $E_p$ of the particles. 
The curve was derived from  Equations~\ref{equ_tc} to \ref{equ_r}, given the detected values $U_{m}=2.45\,\mathrm{mV}$ and $T_c=61\,\mu \mathrm{s}$. The square highlights the case where $E_p=15\,\mathrm{MPa}$ (aerogel), which yields $R=0.98\,\mathrm{mm}$ and 
$v=1.95\,\mathrm{m\,s^{-1}}$. The Young's modulus on a logarithmic scale is shown at the top. See text for details.}
\label{fig_properties}
\end{figure}

Figure~\ref{fig_properties} shows the relationship between $v$, $R$ and $E_p$, given the detected values $U_{m}=2.45\,\mathrm{mV}$ and $T_c=61\,\mu \mathrm{s}$ based on Equations~\ref{equ_tc} to \ref{equ_r}. Estimations of the particle radius and speed are very sensitive to the values of $E_p$. If we simply assume that the detected particle had properties 
equal to those of the aerogel mentioned above, we would conclude that its radius and its impact speed are about $1\,\mathrm{mm}$ and $1.95\,\mathrm{m\,s^{-1}}$ (the square in Figure~\ref{fig_properties} highlights this case).

Nevertheless, in \citet{flandes2013} we thoroughly discussed how the theoretical values deviate 
from the experimental data. We observed that, on average, the signal amplitudes $U_m$ from the 
impacts could be underestimated by up to 30\,\%, while the impact
durations $T_c$ could be overestimated by up to 50\,\% (this can be seen in Table~\ref{table:AE}, 
see also Appendix~2), which 
means that the real $U_m$ values could be larger and the real $T_c$ values could be shorter 
than the recorded ones. Furthermore, our experiments and also the theory show that $T_c$ is 
correlated  with the Young's modulus of the material. This means, 
 for a given impact speed, the more elastic the material, the shorter the impact durations produced and vice versa. A 
 similar effect applies to the voltage amplitude: For a given impact speed, the more elastic the material, the more intense the voltage amplitude and vice versa. Thus, we conclude that for our detected particle the true values should be 
 $U_{m} > 2.45\,\mathrm{mV}$ and $T_c < 61\,\mu s$, and thus, $E_p >15\,\mathrm{MPa}$. Therefore, the derived radius of the particle should be a lower limit, $R>1\,\mathrm{mm}$, and the derived speed should be an upper limit: $v<1.95\,\mathrm{m\,s^{-1}}$, according to Figure~\ref{fig_properties}. Furthermore, the Young's modulus of the nucleus 
surface may be as large as 500~MPa \citep{gibson1982}, which implies an upper limit for the estimated particle 
radius of $2.3\,mm$ and thus, a lower limit for the speed of $0.1\,\mathrm{m\,s^{-1}}$.
{\bf The comparison of the evaluated speed range with speeds measured at 67P for compact 
particle  \citep{dellacorte2015} and for fluffy particles \citep{fulle2015} is in agreement with
a porous nature of the particle.
} 

 Based on our analysis of the $\mathbf{U}_m$ and $\mathbf{T}_c$ values recorded by DIM during Philae's descent 
 and our aerogel experiments (Appendix~2) we conclude that
the detected particle was very likely  
porous, having a porosity of about $40\%$. 

\subsection{Detection threshold and sensitivity range}

\label{sec_sensitivity}

\begin{figure}
   \centering
   \hspace{-3mm}
\includegraphics[width=0.5\textwidth]{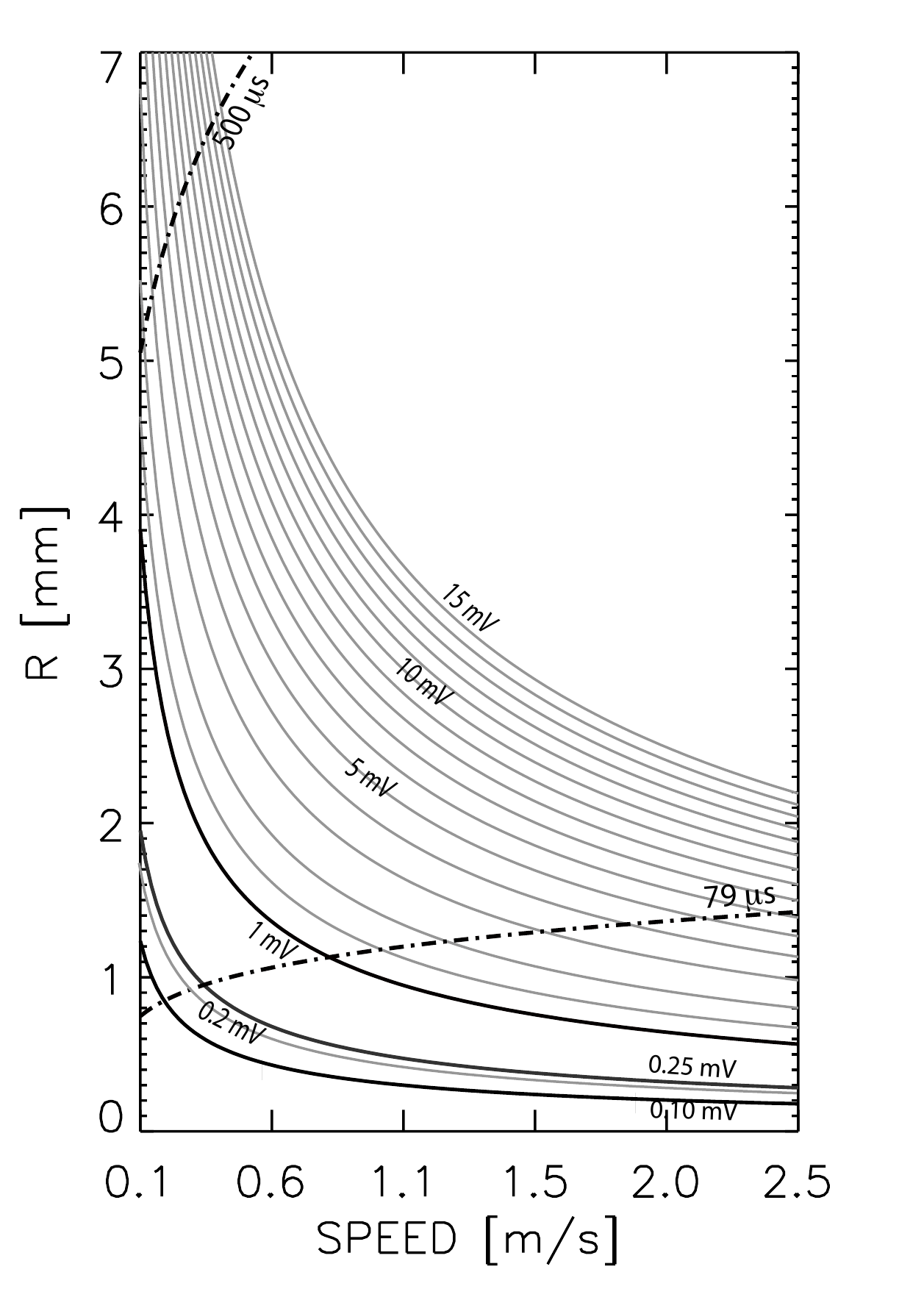}
\caption[]{
DIM's estimated detection range for porous ice particles. The curves are based on the mechanical properties 
of aerogel (bulk density of $250\,\mathrm{kg\,m^{-3}}$, average Young's modulus of $15\,\mathrm{MPa}$ and 
Poisson ratio of $0.2$). Measurement boundaries are specified as dashed lines (see text for details).}
\label{fig_detrange}
\end{figure}

In order to estimate the sensitivity range of DIM, we assume a  voltage range 
$\mathrm{0.2\,mV} < U_m < \mathrm{15\,mV}$. This is shown in Figure~\ref{fig_detrange} where every black solid curve 
represents the particle's radius versus impact speed dependence for a constant signal amplitude $U_m$. 
The lowest possible voltage that can be recorded is determined by the margin (see Section~\ref{sec_margin}) that is
determined by the instrument setup before the measurement. The margin values allowed by the flight software
are $0.1$ ($30\,\mathrm{dB}$), $0.25$ ($40\,\mathrm{dB}$) and $1.0\, \mathrm{mV}$ ($50\,\mathrm{dB}$), respectively, and are represented by continuous thick solid lines in Figure~\ref{fig_detrange}.  
The corresponding impact durations that impacting porous particles could produce range from several tens to hundreds 
of microseconds. Furthermore, the DIM on board electronics can only register impact durations up to a maximum 
value that is determined by the DIM operational mode (see Section~\ref{sec_commanding} and Appendix~1 for details):
the BC mode allows to register impact duration values up to $79\,\mu \mathrm{s}$ and the BCT2 up to $500\,\mu \mathrm{s}$. 
These limits for the impact duration are represented by dashed lines in Figure~\ref{fig_detrange}. 

\section{Discussion}

\label{sec_discussion}

The question arises why DIM did not detect any dust impact at its final landing site Abydos. 
Two possible explanations are obvious: 
(1) Shielding by ambient cliffs close to Philae. Reconstruction of Philae's surroundings from 
the images taken by the CIVA cameras \citep{bibring2015} indicates that Philae 
ended up in some type of cavity. This is also in agreement 
with very low ambient temperatures ranging from 90\,K to 130\,K between night and day \citep{spohn2015}. 
Furthermore, Philae is likely tilted by about 
$90^{\circ}$ w.r.t. the local gravity vector. Reconstruction of the lander's orientation indicates that the +Y sensor 
of DIM may be oriented towards open space while the +X and the +Z sensors
are likely strongly shielded by cliffs that may be only a few meters away. (2) There was 
no cometary activity close to the landing site. The gas spectrometers on board Philae 
detected only a very low gas density \citep{goesmann2015,morse2015}, indicating that the landing site 
and its immediate vicinity were not active when DIM was measuring. Furthermore, due to Philae's operational
constraints at Abydos, most of the DIM measurements were performed at night when 
the comet was likely not active  ({\em cf.} Table~\ref{tab_operation3}).  

The Rosetta Lander Imaging System \citep[ROLIS;][]{mottola2007} detected one single dust particle 
in the vicinity of Philae at Abydos \citep{schroeder2015}. The ROLIS field-of-view is oriented towards 
the bottom of Philae (--Z direction). The particle is seen in front of one of the cliff walls, implying that 
it must have been in the cavity close to Philae. This particle could have been swept up from the 
cometary surface when Philae came to rest at Abydos.
However, the size of the particle cannot easily be determined, and no other particles were detected in 
the cavity. Additional particles are visible in ROLIS and CIVA images, however very likely at much 
larger distances from Philae. 

Many particles can also be seen in images taken by the cameras on board Rosetta. These observations
allowed the detection of particles likely on bound orbits around the comet nucleus and also fast and outflowing
particles that had diameters up to 17~mm were detected \citep{rotundi2015}. 
Dynamical simulations showed that the orbits
of millimeter to centimeter sized particles are relatively unstable while objects
of a few decimeters and larger in diameter may stay in orbit about the cometary nucleus for years, 
i.e. a full orbital revolution of the comet around the Sun \citep{fulle1997}.

The long impact duration measured by DIM during the single particle detection cannot be reproduced 
by impacts of compact particles in laboratory experiments. However, we 
could obtain a very long impact duration in experiments 
with aerogel particles which had a density of $\mathrm{250\,kg\,m^{-3}}$ (see Appendix~2). It 
indicates that the detected particle likely was a porous conglomerate. 

The Cometary Secondary Ion Mass Analyzer (COSIMA) on board 
Rosetta also collected many porous particles in the size range up to a few hundred 
micrometers \citep{schulz2015}. Most of these particles were collected at a distance 
between 10~km and 30~km from the cometary nucleus surface. The particles were very 
fragile so that many of them disintegrated during the collection process. The composition of 
these particles is not known yet, but they were not likely water ice particles.

The motion of particles under the influence of a very dilute gas has been studied by \citet{tal2014} using a Monte Carlo code.  This code is similar to those described in \citet{davidsson2004, davidsson2010}. 
\citet{tal2014} investigated two general scenarios.  The first was an ice patch (''spot``) at some particular temperature, and the second one was a jet.  In the first case the gas flow was uncollimated while in the second case the particle motion was calculated for different degrees of collimation of the gas flow.  She found that for a spot at 155\,K compact ice particles as large as $\mathrm{70\,\mu m}$ could be lifted off from the surface.  The gas drag force on a particle is proportional to its cross section, so for a spherical particle of radius $a$ the drag force can be written as $$F_{drag}=\alpha a^2$$ where $\alpha$ is some constant.  The gravitational force on the  particle is proportional to its mass.  For a spherical particle this means that $$F_{grav}=\beta \rho a^3$$ where $\beta$ is another proportionality constant and $\rho$ is the density of the  particle material.  The maximum liftable 
particle radius will be that for which the gravitational force on the particle exactly equals the drag force.  In this case 
\begin{equation}\label{amax}
a_{max}=\frac{\alpha}{\beta\rho}
\end{equation} and the maximum liftable radius will be inversely proportional to the particle density.

\citet{tal2014} assumed particles with a density 
of 1000\,kg\,m$^{-3}$, corresponding to water ice, but if the material is porous, these results must be adjusted.  The porosity, $p$, is defined as the fraction of the volume that is void of mass.  Thus if a compact body of mass, $m$ occupies a volume, $V$, so that $$\rho=\frac{m}{V}$$ then a body with mass, $m$ and porosity, $p$ will occupy a volume $$V=\frac{m}{(1-p)\rho}=\frac{m}{\rho^*}$$ where $\rho^*=(1-p)\rho$ is the effective density of the porous material.  This is the density that must be used in Eq.\,\ref{amax} so that the maximum liftable radius will be proportional to $((1-p)\rho)^{-1}$.  
These calculations assume that the trajectories of the gas molecules make random angles with the normal to the surface.  However, if the gas is released as a collimated jet, the same gas production rate will provide a more efficient momentum transfer to the particle and somewhat larger particles will be lifted. 

The Visible, Infrared and Thermal 
Imaging Spectrometer (VIRTIS) on board Rosetta measured daytime surface temperatures between 
180 and 230\,K \citep{capaccioni2015}.  
At these temperatures millimeter sized compact particles can easily be lifted off from the surface.  For temperatures as 
low as 170\,K \citet{tal2014} found that sublimation from a spot could lift compact particles of 5.4\,mm radius off 
the surface.  Porous (fluffy) particles could be proportionately larger.  In view of this it is entirely consistent 
that the particle seen by the DIM instrument was lifted off from the comet's surface by sublimating water ice.  
While it is customary to use a total gas production rate instead of a temperature, this can be misleading if what is measured is the average value over a large area.  In such a case local patches of activity can contribute much more gas (and drag) than what would be expected from the average value, so that locally much larger particles can be lifted than would be expected from the averaged gas production rate. 

The lifetime of a particle near the comet depends on its composition and albedo.  Calculations of the lifetime of 
ice particles have been done by \citet{beer2006}. They find that pure ice particles larger than a few microns will survive 
for millions of years at around 3\,AU.  However, adding even a small amount of darker material sharply increases 
the heating rate so that micron sized particles are only expected to survive for hours.  In contrast, millimetre-sized 
particles of dirty ice could survive for many months.  The results of \citet{beer2006} are for compact 
ice particles. The rate of decrease of the radius $a$ of the particle is given by 
\begin{equation}
\frac{da}{dt}=\frac{P_{vap}(T)}{\rho}\sqrt{\frac{m}{2\pi kT}}
\end{equation}
where $P_{vap}$ is the vapor pressure, 
$\rho$ is the density of the particle material, $m$ is the mass of a molecule of particle material, $k$ is Boltzmann's 
constant, and $T$ is the temperature of the particle.  If the particle is porous, the mass density $\rho$ must again be 
replaced by $(1-p)\rho$ and the lifetime of the particle will be shortened proportionately. On the other hand, 
the lifetime of the particle will be greatly increased if it is composed of less volatile material such as 
organics or silicates.

Oblique impacts up to angles $\phi = 50^{\circ}$ were examined in the course of the DIM calibration experiments 
reported by \citet{flandes2013,flandes2014}. Whereas the impact duration $T_c$ increased by 5\%, 
the amplitude decreased by 20\%, both following a cosine law. This entails that the parameter range for $R$ gets 
slightly expanded towards lower values by 10\% if analysed as perpendicular impact.  There are additional effects 
when the impact angles get so large that sliding friction and rolling of the particle sets in \citep{wu2003}. 
These effects further decrease the perpendicular impact force and increase the contact time. 
With only one detected event, it is not possible to take these effects into account.




\section{Conclusions}

\label{sec_conclusions}

Measurements with the Dust Impact Monitor (DIM) on board Philae were obtained 
before the separation from Rosetta, during Philae's descent to the nucleus surface, and 
at Philae's final landing site Abydos. Our results can be summarised as follows:

\begin{itemize}
\item[(1)]DIM detected no dust before the separation of Philae from Rosetta during a measurement period of 
about 4.7~hours. This is most likely due to the
unfavorable detection geometry of the DIM sensor and obscuration by the Rosetta spacecraft 
during this mission phase, preventing the detection
of particles radially ejected from the nucleus. 
\item[(2)] One single dust impact was detected at about 2.4~km altitude during Philae's descent to its nominal
landing site Agilkia. This was very likely a porous particle with radius 
 $ R \approx 1\,\mathrm{mm}$ 
that hit the sensor with an impact speed 
$v \approx 2\,\mathrm{m\,s^{-1}}$.
The particle could have been lifted off from the comet's surface by sublimating water ice.
\item[(3)] At Philae's final landing site Abydos, DIM detected no dust particle during a total measurement
period of 5~hours. This may be due to the strong obscuration of DIM by the ambient cliffs, or the very
low cometary activity at Abydos because most of the measurements were taken at night, or other effects. 
\end{itemize}


The particle detection at 2.4~km altitude is the closest dust detection by a dedicated dust 
detector at 
a cometary nucleus ever. 
In June 2015 contact with Philae was established again, which may provide the 
possibility for additional DIM measurements at Abydos.
The detection of 
more dust particles would allow us to better constrain the small scale  
properties of the cometary nucleus material. 

\begin{acknowledgements}

SESAME is an experiment on the Rosetta lander Philae. It consists of three instruments CASSE, DIM and PP, which 
were provided by a consortium comprising DLR, MPS, FMI, MTA~EK, Fraunhofer IZFP, Univ. Cologne, LATMOS and ESTEC. 
The contribution from MTA~EK to the SESAME-DIM experiment was co-funded through the PRODEX contract No.~90010 and by the Government of Hungary through European Space Agency contracts No 98001, 98072, 4000106879/12/NL/KML and 4000107211/12/NL/KML under the plan for European Cooperating States (PECS).
This research was supported by 
the German Bundesministerium f\"ur Bildung und Forschung through Deutsches
Zentrum f\"ur Luft- und Raumfahrt e.V. (DLR, grant 50\,QP\,1302). Support by MPI f\"ur Sonnensystemforschung
is gratefully acknowledged. A. Flandes was also supported by DGAPA-PAPIIT IA100114. We thank Lorenz Ratke and 
Benjamin Ignatzi (DLR Institut f\"ur Werkstoffforschung K\"oln) for providing 
us with aerogel samples for impact calibration experiments. We are grateful to Cynthia Volkert and 
Cornelia Mewes (Institute for Materials Science, University of G\"ottingen) for supporting us with measurements 
of the Young's modulus of the aerogel particles. We thank the Rosetta project at ESA and the Philae project at 
DLR and CNES for effective and successful mission operations, and the GIADA team from providing us with 
predicted dust fluxes for DIM. We are grateful to an anonymous referee for helpful suggestions concerning the
presentation of our results. 

\end{acknowledgements}


\section*{Appendix 1: Impact parameters}

\label{app_1}

\subsection*{Impact duration}

The conversion between the number of counts $n$ of the 20~MHz clock and the impact duration ${\bf T}_c$ in an
internally defined logarithmic scale is given by: 
\begin{equation}
\mathbf{T}_c = 20 \cdot \log_{10}(2 \cdot n) 
= 20 \cdot \log_{10}\left ( \frac{40\cdot T_c}{10^{-6}\,\mathrm{s}} \right).   \label{eq_dim1}
\end{equation}

The results of Equation~\ref{eq_dim1} are rounded to the nearest integer value. Impacts with $\mathbf{T}_c \geq 70$ are stored as if they were impacts with $\mathbf{T}_c = 70$, thus impact durations larger than $\mathrm{79\,\mu s}$ cannot be distinguished in the BC mode. 

Telemetry of the BCT2 mode additionally includes the counter value $n$. Therefore, this mode delivers impact durations of all accepted impacts with a time resolution of $\mathrm{0.05\,\mu s}$.


The measuring range is limited by the clock frequency (20 MHz corresponding to a resolution of 50 ns; with 6~dB 
bandwidth) and depends on the operation mode. 
Thus, the following instrument limits have been specified: 
$T_{c,min} = 100\,\mathrm{ns}$, 
$T_{c,max,BC} = \mathrm{79\,\mu s}$ for BC mode and 
$T_{c,max,BCT2} = \mathrm{500\,\mu s}$ for BCT2 mode. 

\subsection*{Peak voltage}

The conversion between the signal generated by the piezoelectric sensors and the amplified 
signal is given by the amplifier transfer function of the DIM flight model \citep{peter2002}:
\begin{equation}
U_{out} \mathrm{[V]} = 3.61 + 0.59 \log_{10} \left (\frac{U_{m}}{U_{r,1}}\right ).             \label{equ_transfer}
\end{equation}
where $U_{r,1} = 1\,\mathrm{V}$ is a reference voltage. The logarithmic amplifier was designed such that the 
maximal output voltage is $U_{out,max} = 3\,\mathrm{V}$. Equation~\ref{equ_transfer} was determined by 
calibration procedures.

For storage of the DIM data a decibel scale relative to $U_{r,2} = 5.355\,\mathrm{\mu V}$
is used so that a 3~V output signal of the logarithmic amplifier corresponds to 85.4\,dB in the input signal 
relative to the reference voltage $U_{r,2}$ which is arbitrary and is given by the development history of 
the DIM instrument. Hence,
\begin{equation}
{\bf U}_{out} [dB] = 20 \cdot \log_{10} \left (\frac{U_{m}}{U_{r,2}} \right) = 
20 \cdot \log_{10} \left (\frac{U_{m}}{5.355 \cdot 10^{-6}\,\mathrm{V}}\right).   \label{eq_dim2}
\end{equation}
After some simple mathematics, Equations~\ref{equ_transfer} and~\ref{eq_dim2} can be transferred to 
\begin{equation}
{\bf U}_{out} [dB] = 20 \cdot \log_{10} \left (\frac{U_{m}}{U_{r,2}} \right) 
 = \frac{U_{out} - 0.5}{0.03}.   \label{eq_dim3}
\end{equation}



\section*{Appendix 2: Laboratory experiments with aerogel}

In addition to the calibration experiments reported earlier by \citet{flandes2013,flandes2014}, we performed impact 
experiments with a flight spare sensor of DIM at Max-Planck-Institut f\"ur
Sonnensystemforschung. We used \emph{high-density} aerogel (bulk density $\rho\approx 250\,\mathrm{kg\,m^{-3}}$ and Young's modulus $E_p=15\,\mathrm{MPa}$)
which -- to a first approximation -- is a material whose density and Young's modulus are close to porous ice. 

\begin{figure}[hbtp]
 \centering 
\includegraphics[width=0.49\textwidth]{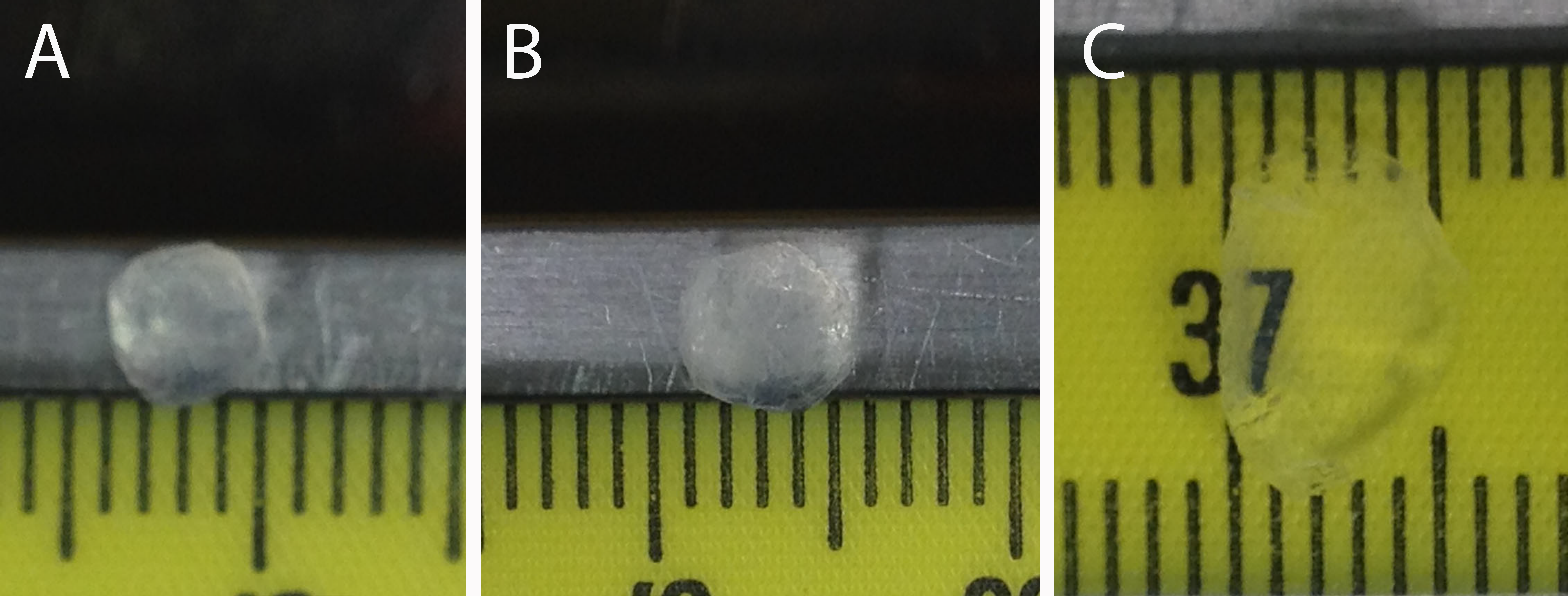}
\caption{Aerogel particles (bulk density, $\rho \approx 250\,\mathrm{kg\,m^{-3}}$) used in the impact experiments reported in Table~\ref{table:AE}.}
\label{fig:AE_TEILE}
\end{figure}

\begin{figure}[hbtp]
 \centering 
 \vspace{-2cm}
\includegraphics[width=.53\textwidth]{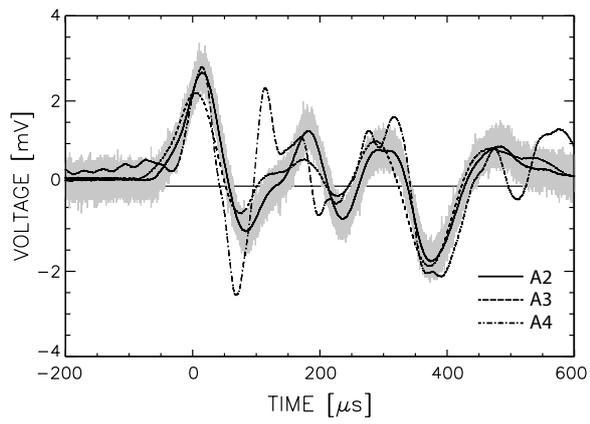}
\vspace{-4.5cm}
\caption{Impact signals generated by the semi-spherical aerogel particle A (Figure~\ref{fig:AE_TEILE}A and Table~\ref{table:AE}, experiments A2 to A4). All curves except the grey one are 
smoothed in order to remove noise. The solid black curve (A2) and the grey curves are the same. The grey curve illustrates the 
noise of each signal. }
\label{fig:AEROG_signal}
\end{figure}

The three aerogel particles shown in Figure~\ref{fig:AE_TEILE} were released from a height of $4.5\,\mathrm{cm}$ (which corresponds 
to an impact speed of approximately $0.94\,\mathrm{m\,s^{-1}}$) onto the centre of the third PZT of the Z-side of a DIM flight 
spare sensor (normal impacts). These particles produced low impact amplitudes of a few millivolts as well as long contact 
durations of many tens of microseconds. These values are consistent with those recorded by the onboard DIM electronics during our
particle impact. The values obtained during our impact experiments are shown in Table~\ref{table:AE} and the signals generated by 
the almost spherical particle (Figure~\ref{fig:AE_TEILE}A) from the oscilloscope data are shown in Figure~\ref{fig:AEROG_signal}. 
In \citet{flandes2013,flandes2014} we discussed in detail how the theoretical values deviate from the 
experimental data. 

\section*{Appendix 3: Error estimation}

Columns (5) and (6) of Table~\ref{table:AE} show the theoretically estimated particle diameter, $D_t$, 
and impact speed, $v_t$, calculated with Equations~\ref{equ_v} and ~\ref{equ_r} and with the measured signal amplitude 
$U_m$ and impact duration $T_c$ (third and fourth columns in Table~\ref{table:AE}). 
The measured values 
do not represent a rigorous statistical analysis, but they show
that the detected particle could be as porous as aerogel, having a density and Young's modulus of the same 
order as aerogel. 
The average estimated uncertainties of these values with respect to the measured signal amplitudes and impact durations 
(columns 3 to 4), are about 40\% for the diameter and 50\% for the speed. Furthermore, so far, our analysis 
assumes that the impacting particles are spherical (see Section~\ref{sec_properties}), which is certainly only
a rough approximation. The shape plays a significant role as well, as can be seen in Table~\ref{table:AE}.

\bibliography{pape,references}

\clearpage



\begin{table}[th]
\caption{DIM measurement details at comet 67P when Philae was connected to 
Rosetta. Column~1 lists the measurement number, column~2 and 3 give the start time of 
each measurement block (all times are given 
in lander on board time), column~4 lists the distance from the comet barycenter,
and column~5 gives the margin. 
All measurements were collected with the X sensor side 
in BCT2 mode with a measurement duration of 600~s. DIM was partially obscured by Rosetta 
during these measurements, see text for details. 
\label{tab_operation}
}
\vskip4mm
\centering
\begin{tabular}{ccccc}
\hline
Meas. & Day        & Start & Distance & Margin   \\
Number&           & Time  &          &          \\
      &    &$\mathrm{[UTC]}$ & [km]  &[dB]          \\
    (1)    &  (2)  & (3)  &  (4)   &  (5)\\
\hline
1 & 16.10.2014 & 20:12:03 & 11.8 & 30       \\
2 & 16.10.2014 & 20:22:23 & 11.8 & 40       \\
3 & 16.10.2014 & 20:36:04 & 11.8 & 30       \\
4 & 16.10.2014 & 20:46:24 & 11.8 & 40       \\
&&&&\\[-1.5ex]
5 & 17.10.2014 & 08:42:04 & 11.0 & 30       \\
6 & 17.10.2014 & 08:52:24 & 11.0 & 40       \\
7 & 17.10.2014 & 09:06:04 & 10.9 & 30       \\
8 & 17.10.2014 & 09:16:25 & 10.9 & 40       \\
&&&&\\[-1.5ex]
9 & 17.10.2014 & 10:32:06 & 10.7 & 30       \\
10 & 17.10.2014 & 10:42:26 & 10.7 & 40       \\
11 & 17.10.2014 & 10:56:06 & 10.7 & 30       \\
12 & 17.10.2014 & 11:06:26 & 10.6 & 40       \\
&&&&\\[-1.5ex]
13 & 17.10.2014 & 12:42:06 & 10.4 & 30       \\
14 & 17.10.2014 & 12:52:26 & 10.4 & 40       \\
15 & 17.10.2014 & 13:06:06 & 10.3 & 30       \\
16 & 17.10.2014 & 13:16:26 & 10.3 & 40       \\
&&&&\\[-1.5ex]
17 & 17.10.2014 & 13:52:06 & 10.2 & 30       \\
18 & 17.10.2014 & 14:02:26 & 10.2 & 40       \\
19 & 17.10.2014 & 14:16:06 & 10.2 & 30       \\
20 & 17.10.2014 & 14:26:26 & 10.1 & 40       \\
&&&&\\[-1.5ex]
21& 17.10.2014 & 15:02:04 & 10.0 & 30       \\
22& 17.10.2014 & 15:12:25 & 10.0 & 40       \\
23& 17.10.2014 & 15:26:04 & 10.0 & 30       \\
24& 17.10.2014 & 15:36:25 &  9.9 & 40       \\
&&&&\\[-1.5ex]
25& 17.10.2014 & 17:22:06 &  9.6 & 30       \\
26& 17.10.2014 & 17:32:26 &  9.6 & 40       \\
27& 17.10.2014 & 17:46:06 &  9.6 & 30       \\
28& 17.10.2014 & 17:56:26 &  9.6 & 40       \\
\hline
\end{tabular}
\end{table}

\clearpage

\begin{table}[th]
\caption{DIM measurement details on 12 November 2014 when Philae was still connected to 
Rosetta and during the descent to the comet's surface. Column~1 lists the measurement number, 
columns ~2 and 3 give the start 
time of each measurement block, column~4 gives the distance from the
barycenter of the comet nucleus, column~5 lists the measurement time, column~6 gives
the sensor side operated, column~7 lists the margin, column~8 lists the fraction 
of the signal-free time in the measurement period, column~9  lists the frequency of false 
signals,
and column~10 gives some comments.  All measurements were collected in BCT2 mode. 
\label{tab_operation2}
}
\vskip4mm
\centering
\begin{tabular}{cccccccccl}
\hline
Meas. & Day        & Start & Dist. & Meas.& DIM  & Margin &  Signal-free & False signal     & \multicolumn{1}{c}{Comments} \\
Numb.&    & Time  &       & Time & Side &        &  period     & Frequency &    \\
      &   &$\mathrm{[UTC]}$&[km]  & [s]&      &  [dB]  &   [\%]     &$\mathrm{[s^{-1}]}$ &  \\
    (1)    &  (2)  & (3)  &  (4) & (5)    &  (6) &   (7)      &   (8)     & (9)   &  \multicolumn{1}{c}{(10)}\\
\hline
29 & 12.11.2014 & 07:38:02 & 25.3 & 100 & X  &  40     & 100        &    --     & DIM partially obscured by Rosetta\\
&&&&&&&&& \\[-1ex]
-- & 12.11.2014 & 08:35:00 & 22.4 & --  & -- &  --    &  --         &  --       & Philae release from Rosetta \\[-1ex]
&&&&&&&&& \\
30 & 12.11.2014 & 08:38:32 & 22.2 & 200 & X  &  40      &   0        &    --     & Meas. terminated at 08:38:45~UTC\\
31 & 12.11.2014 & 08:42:23 & 22.1 & 200 & Y  &  40      &   0        &    83     & \\
32 & 12.11.2014 & 08:46:13 & 22.0 & 200 & Z  &  40      &  99        &    68     & \\
&&&&&&&&& \\[-1.5ex]
33 & 12.11.2014 & 08:50:03 & 21.8 & 200 & X  &  50      & 100        &    --     & \\
34 & 12.11.2014 & 08:53:52 & 21.6 & 200 & Y  &  50      & 100        &    --     & \\
35 & 12.11.2014 & 08:57:42 & 21.4 & 200 & Z  &  50      & 100        &    --     & \\
&&&&&&&&& \\[-1.5ex]
36 & 12.11.2014 & 09:59:04 & 18.6 & 200 & X  &  40      &   0        &   112     & \\
37 & 12.11.2014 & 10:02:54 & 18.4 & 200 & Y  &  40      &  99        &    32     & \\
38 & 12.11.2014 & 10:06:44 & 18.3 & 200 & Z  &  40      &   0        &    77     & \\
&&&&&&&&& \\[-1.5ex]
39 & 12.11.2014 & 14:40:04 & 5.1 & 100 & X  &  40      &  96        &    93     & \\
40 & 12.11.2014 & 14:42:14 & 5.0 & 100 & Y  &  40      &  98        &    47     & Noise terminated at 14:42:26~UTC \\
41 & 12.11.2014 & 14:44:24 & 4.9 & 100 & Z  &  40      &   0        &    77     & \\
&&&&&&&&& \\[-1ex]
-- & 12.11.2014 & 15:34:04 & 2.3 & --  & -- & --       & --         &  --       & Philae touchdown at Agilkia  \\
\hline
\end{tabular}
\end{table}

\clearpage

\begin{table}[th]
\caption{DIM measurement details when Philae was at its final landing site Abydos. 
Column~1 lists the measurement number, column~2 and 3 give the start time of each measurement block,
column~4 gives the measurement duration, column~5 the sensor side operated,
column~6 lists the margin, and 
column~7 gives Philae's solar panel that was illuminated during the measurement. 
All measurements were taken in BC mode.
\label{tab_operation3}
}
\vskip4mm
\centering
\begin{tabular}{ccccccc}
\hline
Meas. & Day        & Start & Meas.& DIM  &  Margin &   Illuminated  \\
Number&    &       & Time & Side &         &   solar array   \\
      &   &$\mathrm{[UTC]}$&[s]&     &  [dB]   &        \\
 (1)   &  (2)  & (3)  &  (4) & (5)   &   (6) & \multicolumn{1}{c}{(7)}\\
\hline
42 &13.11.2014 & 07:02:22 & 558 & X  &  40     &  2  \\
43 &13.11.2014 & 07:12:10 & 558 & Y  &  40     &  None           \\
44 &13.11.2014 & 07:21:59 & 558 & Z  &  40     &  None           \\
&&&&&&\\[-1.5ex]
45 &13.11.2014 & 07:33:22 & 558 & X  &  30     &  None              \\
46 &13.11.2014 & 07:43:21 & 558 & Y  &  30     &  None              \\
47 &13.11.2014 & 07:53:08 & 558 & Z  &  40     &  None              \\
&&&&&&\\[-1.5ex]
48 &13.11.2014 & 09:04:24 & 558 & X  &  40     &  None          \\
49 &13.11.2014 & 09:14:13 & 558 & Y  &  40     &  None          \\
50 &13.11.2014 & 09:24:02 & 558 & Z  &  40     &  None          \\
&&&&&&\\[-1.5ex]
51 &13.11.2014 & 09:35:24 & 558 & X  &  40     &  None          \\
52 &13.11.2014 & 09:45:12 & 558 & Y  &  40     &  None          \\
53 &13.11.2014 & 09:55:01 & 558 & Z  &  40     &  None          \\
&&&&&&\\[-1.5ex]
54 &13.11.2014 & 11:06:24 & 558 & X  &  40     &  None          \\
55 &13.11.2014 & 11:16:13 & 558 & Y  &  40     &  None          \\
56 &13.11.2014 & 11:26:02 & 558 & Z  &  40     &  None          \\
&&&&&&\\[-1.5ex]
57 &13.11.2014 & 11:37:24 & 558 & X  &  40     &  None          \\
58 &13.11.2014 & 11:47:13 & 558 & Y  &  40     &  None          \\
59 &13.11.2014 & 11:57:02 & 558 & Z  &  40     &  None          \\
&&&&&&\\[-1.5ex]
60 &13.11.2014 & 13:08:25 & 558 & X  &  40     &  None          \\
61 &13.11.2014 & 13:18:14 & 558 & Y  &  40     &  None          \\
62 &13.11.2014 & 13:28:03 & 558 & Z  &  40     &  None          \\
&&&&&&\\[-1.5ex]
63 &13.11.2014 & 13:39:25 & 558 & X  &  40     &  None          \\
64 &13.11.2014 & 13:49:14 & 558 & Y  &  40     &  None          \\
65 &13.11.2014 & 13:59:03 & 558 & Z  &  40     &  None          \\
&&&&&&\\[-1.5ex]
66 &14.11.2014 & 06:33:42 & 557 & X  &  30     &  2              \\
67 &14.11.2014 & 06:43:30 & 557 & Y  &  40     &  2              \\
68 &14.11.2014 & 06:53:18 & 557 & Z  &  30     &  2              \\
&&&&&&\\[-1.5ex]
69 &14.11.2014 & 07:04:41 & 558 & X  &  30     &  2/3        \\
70 &14.11.2014 & 07:14:30 & 558 & Y  &  40     &  2/3        \\
71 &14.11.2014 & 07:24:19 & 558 & Z  &  40     &  2/3        \\
\hline
\end{tabular}
\end{table}

\clearpage

\begin{table}[hbtp]
\caption{Average properties$^{\dagger}$ of the PZTs and the particle.}
\centering
\small
\begin{tabular}{lccc}
\hline\\[-1.8ex]
      Property                &Symbol       &Measured value                         \\[0.8ex] 
\hline\\[-1.8ex]
	Piezoelectric constant   &	$d_{33}$	   &	$\mathrm{239 \cdot 10^{-12}\,m\,V^{-1}}$    \\
	Capacitance$$     &  $C$        &	$\mathrm{14.2\,nF}$	                        \\ [0.8ex]
	PZT's reduced Young's modulus             &  $E_{PZT}$	       &	$\mathrm{74.5\,GPa}$\\
    Particle's reduced Young's modulus             &	$E_p$      &	$\mathrm{>15\,MPa}$  \\
    Particle's bulk density     &  $\rho$        &	 $\mathrm{\gtrsim250\,{kg\,m^{-3}}}$   \\ [0.8ex]
\hline
\multicolumn{3}{l}{}
$^{\dagger}$ $E_p$ and $\rho$ were measured for aerogel (See Section~\ref{sec_properties} and Appendix~2). 
\end{tabular}
\label{table:PIEZO}
\end{table}


\begin{table}[hbtp]
\caption{Measured voltage amplitude, $U_m$, and impact duration, $T_c$, for impacts of aerogel particles. 
 $D$ (column~2) is the diameter of 
the particle as shown in Figure~\ref{fig:AE_TEILE}. $D_t$ (column~5) and $v_t$ (column~6) are the diameter and speed 
derived from Equations~\ref{equ_v} and ~\ref{equ_r} based on the measured amplitude $U_m$ (column~3) and impact 
duration $T_c$ (column~4). We assumed a density $\rho=250\,\mathrm{kg\,m^{-3}}$ and Young's modulus $E_p=15\,\mathrm{MPa}$.
The measurements were obtained with an impact speed of about $0.94\,\mathrm{m\,s^{-1}}$ on the Z-side (3rd PZT) of DIM.}
\centering
\small
\begin{tabular}{{l}{c}{c}{c}{c}{c}{c}}
\hline\\[-1.4ex]
No.      &$D$ [mm]      &$U_m$ [mV]   & $T_c\,[\mathrm{\mu s}]$ &  $D_t$ [mm]  &$v_t$ [$\mathrm{m\,s^{-1}}$] & Notes      \\[0.8ex] 
   (1)   &      (2)     &    (3)      &(4)                      & (5)          &(6)         & (7)         \\ 
\hline\\[-1.8ex]	
	C1   &5-7&	$1.41$	   &	$64.45$ & $5.03$  & $1.16$  & Irregular, Fig.~\ref{fig:AE_TEILE}C    \\
	C2   &	&	$6.06$	   &	$58.79$ & $5.70$  & $3.23$  &	    \\
\hline
	B1   &5&	$2.18$	   &	$77.40$ & $5.96$  & $1.21$  & Irregular/spherical, Fig.~\ref{fig:AE_TEILE}B    \\
	B2   & &	$3.17$	   &	$76.01$ & $6.17$  & $1.56$  &		    \\
\hline
	A1   &4&	$3.40$	   &	$68.54$ & $5.84$  & $1.86$  & Semi-spherical, Fig.~\ref{fig:AE_TEILE}A    \\	
	A2   & &	$3.67$	   &	$87.78$ & $6.88$  & $1.43$  &           \\
	A3   & &	$2.11$	   &	$92.74$ & $6.65$  & $0.95$  &	    \\	
	A4   & &	$3.74$	   &	$74.48$ & $6.22$  & $1.78$  &	    \\
\hline			
\multicolumn{3}{l}{}
\end{tabular}
\label{table:AE}
\end{table}

\clearpage




\end{document}